\documentclass{aa}   
\pdfoutput=1   
\usepackage{txfonts} 
\usepackage{amssymb}
\usepackage{epstopdf}
\usepackage{epsfig}
\usepackage{natbib}
\bibpunct{(}{)}{;}{a}{}{,} 

\title{A study of deuterated water in the low-mass protostar IRAS16293-2422 \thanks{based on {\it Herschel}/HIFI observations. {\it Herschel} is an ESA space observatory with science instruments provided by European-led Principal Investigator consortia and with important participation from NASA.}}

\author{Coutens A. \inst{\ref{cesr},\ref{cesr2}} \and
Vastel C.  \inst{\ref{cesr},\ref{cesr2}} \and
Caux E. \inst{\ref{cesr},\ref{cesr2}} \and
Ceccarelli C. \inst{\ref{ipag}}  \and
Bottinelli S. \inst{\ref{cesr},\ref{cesr2}} \and
Wiesenfeld L.  \inst{\ref{ipag}}  \and
Faure A.  \inst{\ref{ipag}} \and
Scribano Y. \inst{\ref{bourgogne}} \and
Kahane C. \inst{\ref{ipag}}
          }

\institute{
Universit\'{e} de Toulouse, UPS-OMP, IRAP, Toulouse, France, \email{audrey.coutens@irap.omp.eu}
\label{cesr}
\and CNRS, IRAP, 9 Av. Colonel Roche, BP 44346, F-31028 Toulouse Cedex 4, France
\label{cesr2}
\and Institut de Plan\'{e}tologie et d'Astrophysique de Grenoble (IPAG), UMR 5274, UJF-Grenoble 1/CNRS, F-38041 Grenoble, France
\label{ipag}
\and Laboratoire Interdisciplinaire Carnot de Bourgogne, UMR 5209-CNRS, 9 Av. Alain Savary, BP47870, F-21078 Dijon Cedex, France
\label{bourgogne}
          }

\date{Received 4 July 2011 / Accepted 6 January 2012}              

\abstract 
{Water is a primordial species in the emergence of life, and comets may have brought a large fraction to Earth to form the oceans. To understand the evolution of water from the first stages of star formation to the formation of planets and comets, the HDO/H$_2$O ratio is a powerful diagnostic. }
{Our aim is to determine precisely the abundance distribution of HDO towards the low-mass protostar IRAS16293-2422 and learn more about the water formation mechanisms by determining the HDO/H$_2$O abundance ratio.}
{A spectral survey of the source IRAS16293-2422 was carried out in the framework of the CHESS (Chemical HErschel Surveys of Star forming regions) Herschel Key program with the HIFI  (Heterodyne Instrument for the Far-Infrared) instrument, allowing detection of numerous HDO lines. Other transitions have been observed previously with ground-based telescopes. The spherical Monte Carlo radiative transfer code RATRAN was used to reproduce the observed line profiles of HDO by assuming an abundance jump. To determine the H$_2$O abundance throughout the envelope, a similar study was made of the H$_2^{18}$O observed lines, as the H$_2$O main isotope lines are contaminated by the outflows.} 
{It is the first time that so many HDO and H$_2^{18}$O transitions have been detected towards the same source with high spectral resolution. We derive an inner HDO abundance ($T$ $\ge$ 100\,K) of about  1.7 $\times$ 10$^{-7}$ and an outer HDO abundance ($T$ $<$ 100\,K) of about 8 $\times$ 10$^{-11}$. To reproduce the HDO absorption lines observed at 894 and 465 GHz, it is necessary to add an absorbing layer in front of the envelope. It may correspond to a water-rich layer created by the photodesorption of the ices at the edges of the molecular cloud. 
At a 3$\sigma$ uncertainty, the HDO/H$_2$O ratio is 1.4-5.8\% in the hot corino, whereas it is 0.2-2.2\% in the outer envelope. It is estimated at $\sim$4.8\% in the added absorbing layer.}
 { Although it is clearly higher than the cosmic D/H abundance, the HDO/H$_2$O ratio remains lower than the D/H ratio derived for other deuterated molecules observed in the same source.
The similarity of the ratios derived in the hot corino and in the added absorbing layer suggests that water formed before the gravitational collapse of the protostar, contrary to formaldehyde and methanol, which formed later once the CO molecules had depleted on the grains.
}


\begin{document}
\maketitle

\abstract

\section{Introduction}

Water is one of the most important molecules in the solar system and beyond, but an unresolved question remains of how water has evolved from cold prestellar cores to protoplanetary disks and consequently oceans for the Earth's specific, but probably not isolated, case. In addition to being a primordial ingredient in the emergence of life, this molecule plays an essential role in the process of star formation through the cooling of warm gas. It also controls the chemistry for many species, either in the gas phase or on the grain surfaces.
In cold dense cores, the gas-phase water abundance is low, less than about $10^{-8}$ \citep[see for example][]{Bergin2002,Caselli2010}, while it can reach much higher values in the outflowing gas of the low-mass protostars environments \citep[$\sim$10$^{-5}$-10$^{-4}$;][]{Liseau1996,Lefloch2010,Kristensen2010} before the solar-type protoplanetary systems are formed. 

In standard gas-phase chemistry, H$_2$O forms through ion-molecule reactions that leads to H$_3$O$^+$, which can dissociatively recombine to form H$_2$O \citep[e.g.,][]{Bates1986,Rodgers2002}, or through the highly endothermic reaction O + H$_2$ $\rightarrow$ OH + H, followed
by the reaction of OH with H$_2$  \citep{Wagner1987,Hollenbach1989,Atkinson2004}. The former process is typical of diffuse cloud conditions, whereas the latter only works in regions with high temperatures, such as shocks or hot cores. It has been realized for around 30 years that, in cold and dense regions, H$_2$O is formed much more efficiently on the grains through a series of reactions involving O and H accreted from the gas \citep[e.g.,][]{Tielens1982,Jones1984,Mokrane2009,Dulieu2010,Cuppen2010}. Near protostars, the grain temperature rises above $\sim$100\,K, leading to a fast H$_2$O ice desorption \citep{Ceccarelli1996,Fraser2001} that increases the H$_2$O gas-phase abundance in the inner part of the envelope called hot core/corino for a high-mass/low-mass protostar \citep{Melnick2000,Ceccarelli1999,Ceccarelli2000}.

Deuterated water is likely to form with a similar chemistry to that of water. 
Theoretically, the HDO/H$_2$O ratio should be high if water has been formed at low temperature, i.e. on cold grain surfaces, and low if it is a product of the photodissociation region or shock chemistry.
Indeed, this is caused by the zero-point energy difference in the vibrational potential \citep{Solomon1973}. As the deuterated species have higher reduced mass than their undeuterated counterparts, the zero-point energy of the deuterated species is lower \citep[the difference in energy between H$_2$O and HDO is 886 K; ][]{Hewitt2005}. Consequently, the enrichment of deuterated water with respect to its main isotopologue takes place at low temperature.
The deuteration fraction observed in high-mass hot cores is typically HDO/H$_2$O $\le 10^{-3}$ \citep{Jacq1990,Gensheimer1996,Helmich1996}, although higher values ($\sim$10$^{-2}$) have recently been found for Orion \citep{Persson2007,Bergin2010}. This ratio has also been estimated in the inner envelope of low-mass protostars such as IRAS16293-2422 \citep{Parise2005} at about 3\%, NGC1333-IRAS4B \citep{Jorgensen2010} with an upper limit of $0.06\%$, and NGC1333-IRAS2A \citep{Liu2011} with a lower limit of  1\%.
Deuterated water has also been sought in ices towards several protostars, and has allowed \citet{Dartois2003} and \citet{Parise2003} to obtain upper limits of solid HDO/H$_2$O from 0.5\% to 2\%.

It is important to know the HDO/H$_2$O ratio throughout the protostar envelope in order to determine how this ratio is preserved after the dispersion of the envelope, when a protoplanetary disk is left 
over, from which asteroids, comets, and planets may form. Obviously, this ratio is needed in order to evaluate the contribution of comets for transferring water in Earth's oceans \citep{Morbidelli2000,Raymond2004,Villanueva2009}, directly associated with the emergence of life.
It seems therefore crucial to determine how similar the observed HDO/H$_2$O ratios in protostellar environments are to those observed in comets and in ocean water on Earth \citep[$\sim$ 0.02 \%, e.g.,][]{Bockelee1998,Lecuyer1998}. Recently, \citet{Hartogh2011} have reported a D/H ratio in the Jupiter family comet 103P/Hartley2, originating in the Kuiper belt, of 0.016, suggesting that some of Earth's water comes from the same comet family.

IRAS16293-2422 (hereafter IRAS16293) is a solar-type protostar situated in the LDN 1689N cloud in Ophiuchus at a distance of 120 pc \citep{Knude1998,Loinard2008}.
It is constituted of two cores IRAS16293A and IRAS16293B separated by $\sim$5$\arcsec$, and the source IRAS16293A itself could be a binary system \citep{Wootten1989}. Several outflows have also been detected in this source \citep{Castets2001,Stark2004,Chandler2005,Yeh2008}.
This Class 0 protostar is the first source where a hot corino has been discovered \citep{Ceccarelli2000b,Cazaux2003,Bottinelli2004}.
It is also a well-studied case thanks to its high deuterium fractionation. For example, the abundance of doubly deuterated formaldehyde has been estimated between 3 and 16$\%$ of the main isotopologue  abundance \citep{Ceccarelli1998,Ceccarelli2001}. Methanol also shows a high deuterium fractionation: about 30 $\pm$ 20\% for CH$_2$DOH, 6 $\pm$ 5\% for CHD$_2$OH, and $\sim$1.4\% for CD$_3$OH \citep{Parise2004}. More recently, \citet{Demyk2010} have determined a methyl formate deuterium fractionation of $\sim$15\%, and  \citet{Bacmann2010} have concluded that there is a ND/NH ratio between 30\% and 100\%.
Singly deuterated water in IRAS16293 has been studied with ground-based telescopes by \citet{Stark2004} and \citet{Parise2005}.
The former find a constant abundance of 3 $\times$ 10$^{-10}$ throughout the envelope with the JCMT observation of the HDO 1$_{0,1}$-0$_{0,0}$ fundamental line at 465 GHz alone,
whereas the latter obtain an inner abundance (where T $\ge$ 100\,K) $X_{\rm in}$ = 1 $\times$ 10$^{-7}$ and an outer abundance $X_{\rm out}$ $\le$ 1 $\times$ 10$^{-9}$ using four transitions observed with the IRAM\footnote{Institut de RadioAstronomie Millim\'etrique}-30m telescope, as well as a JCMT\footnote{James Clerk Maxwell Telescope} observation at 465 GHz.
 Using the water abundances determined from ISO/LWS\footnote{Infrared Space Observatory/Long Wavelength Spectrometer} observations, which suffer from both a high beam-dilution and very low spectral resolution \citep{Ceccarelli2000}, \citet{Parise2005} estimated a deuteration ratio HDO/H$_2$O of about 3$\%$ in the hot corino and lower than 0.2$\%$ in the outer envelope. 
 Thanks to the spectral survey carried out with Herschel/HIFI towards IRAS16293 in the framework of the CHESS key program \citep{Ceccarelli2010}, numerous HDO and H$_2^{18}$O transitions have been observed at high spectral resolution, allowing an accurate determination of the HDO/H$_2$O ratio in this source.
 Heavy water (D$_2$O) has also been detected in IRAS16293 with the 1$_{1,1}$-0$_{0,0}$ fundamental ortho transition at 607 GHz with Herschel/HIFI  \citep{Vastel2010} and the 1$_{1,0}$-1$_{0,1}$ fundamental para transition at 317 GHz with JCMT  \citep{Butner2007}. From both transitions, \citet{Vastel2010} estimated a D$_2$O abundance of about 2 $\times$ 10$^{-11}$ in the colder envelope.
 
The main goal of this paper is to determine the abundance of HDO throughout the protostar envelope, using new HDO collisional coefficients with H$_2$ computed by \citet{Faure2011} and \citet{Wiesenfeld2011}, and combining the Herschel/HIFI, JCMT and IRAM data in this source.
In Sect. 2, we present the observations and in Sect. 3, the modeling of the HDO and H$_2^{18}$O emission distribution in the source using the radiative transfer code RATRAN \citep{Ratran}. In Sect. 4, we discuss the derived water deuterium fractionation, and conclude in Sect. 5.

\section{Observations}

In total, thirteen HDO transitions have been detected towards the solar-type protostar IRAS16293, nine with the Herschel HIFI instrument, three with the IRAM-30m, and one with the JCMT. Three other transitions observed with HIFI, although not detected, have been used to derive upper limits. In addition, fifteen transitions of H$_2^{18}$O have been observed with HIFI.  The ortho-H$_2^{18}$O 1$_{1,0}$-1$_{0,1}$ fundamental line is clearly detected, four other transitions are tentatively detected, and ten others were not detected but their upper limits give constraints on the water abundance. Numerous H$_2^{16}$O lines have been detected but not used since contaminated by the outflows. The ortho-H$_2^{17}$O 1$_{1,0}$-1$_{0,1}$ fundamental line has also been detected.

\subsection{HIFI data}

In the framework of the guaranteed time Key Program CHESS \citep{Ceccarelli2010}, we observed the low-mass protostar IRAS16293 with the HIFI instrument  \citep{deGraauw2010} onboard the Herschel Space Observatory \citep{Pilbratt2010}. 
A full spectral coverage was performed in the frequency ranges [480 -- 1142 GHz] (HIFI bands 1 to 5), [1481 -- 1510 GHz] (HIFI band 6a), and [1573 -- 1798 GHz] (HIFI bands 6b and 7a).
The observations were obtained in March 2010 and February 2011, using the HIFI Spectral Scan Double Beam Switch (DBS) fast-chop mode with optimization of  the continuum. 
Twelve HDO lines and fifteen H$_2^{18}$O lines have been observed in these bands. Table \ref{obs} lists, for all transitions, the observed parameters.
The HIFI Wide Band Spectrometer (WBS) was used, providing a spectral resolution of 1.1 MHz 
(0.69\,km\,s$^{-1}$ at 490 GHz and 0.18\,km\,s$^{-1}$ at 1800 GHz) 
over an instantaneous bandwidth of 4$\times$1\,GHz. 
The targeted coordinates were $\alpha_{2000}$ = 16$^h$ 32$^m$ 22$\fs$75, $\delta_{2000}$ = $-$ 24$\degr$ 28$\arcmin$ 34.2$\arcsec$, a position at equal distance of IRAS16293 A and B, to measure the emission of both components. 
The DBS reference positions were situated approximately 3$\arcmin$ east and west of the source. 
The forward efficiency is about 0.96 at all frequencies. 
The main beam efficiencies used are shown in Table \ref{obs} and are the recommended values from \citet{Roelfsema2012}. 

The data were processed using the standard HIFI pipeline up to frequency and amplitude calibrations (level 2) with the ESA-supported package HIPE 5.1 \citep{ott2010} for all the bands except band 3a, processed with the package HIPE 5.2. 
In the selected observing mode, all lines were observed at least four times (if they are on a receiver band edge), but generally eight times (four in LSB and four in USB) for each polarization.
To produce the final spectra, all observations were exported to the GILDAS/CLASS\footnote{http://www.iram.fr/IRAMFR/GILDAS/} software. Using this package, the H and V polarizations at the line frequencies were averaged, weighting them by the observed noise for each spectra. We verified, for all spectra, that no emission from other species was present in the image band.
HIFI operates as a double sideband (DSB) receiver, and the gains for the upper and lower sidebands are not necessarily equal. From the in-orbit performances of the instrument \citep{Roelfsema2012},  a sideband ratio of unity is assumed for the HDO transition seen in absorption against the continuum (band 3b). The observed continuum is therefore divided by two to obtain the SSB (single sideband) continuum. The relative calibration budget error of the HIFI instrument is presented in Table 7 of \citet{Roelfsema2012}.
Considering all the upper limits and estimated errors, we assumed an overall calibration uncertainty of 15\% for each line.

\subsection{IRAM data}

The IRAS16293 Millimeter And Submillimeter Spectral Survey \citep[TIMASSS;] []{Caux2011} was performed between 2004 and 2007 at the IRAM-30m telescope in the frequency range 80-280 GHz and at the JCMT-15m telescope in the frequency range 328-368 GHz. The observations were centered on the IRAS16293 ÒBÓ source at $\alpha_{2000}$ = 16$^h$ 32$^m$ 22$\fs$6, $\delta_{2000}$= $-$ 24$\degr$ 28$\arcmin$ 33$\arcsec$.
In this survey, four HDO lines were detected, but only three of them (80.578, 225.897, and 241.561 GHz) have been used in this paper, as the fourth transition at 266.161 GHz lies in a part of the survey where the calibration uncertainty is very high \citep{Caux2011}.
The spectral resolution is 0.31 MHz ($\sim$1.2\,km\,s$^{-1}$) at 81 GHz and 1 MHz ($\sim$1.3\,km\,s$^{-1}$) at 226 GHz and 242 GHz.

\subsection{JCMT data}
The HDO 1$_{0,1}$-0$_{0,0}$ fundamental transition at 464.924 GHz was previously observed by \citet{Stark2004} on the IRAS16293 A source ($\alpha_{2000}$ = 16$^h$ 32$^m$ 22$\fs$85, $\delta_{2000}$= $-$ 24$\degr$ 28$\arcmin$ 35.5$\arcsec$) and by \citet{Parise2005} on the IRAS16293 B source. The JCMT beam at this frequency is about 11\arcsec. 
Despite the different pointings, the profiles and the intensities of the line obtained are similar. This line shows both a wide emission and a narrow deep absorption (see Fig. \ref{comp_abs}).
The observations are not optimized for an accurate continuum determination. However, the continuum level is a crucial parameter in the following modeling. 
We see thereafter that we can estimate it at about 1 K thanks to the modeling and the dust emissivity model we chose. It consequently means that the narrow self-absorption completely absorbs the continuum.

\begin{table*}[!ht]
\begin{center}
\caption{Parameters for the observed HDO, H$_2^{18}$O, H$_2^{17}$O, and HD$^{18}$O lines$^{(1)}$.}
\label{obs}
\begin{tabular}{ l r c r c c c c c r c c c}
\hline
\hline	
Species & Frequency & J$_{\rm Ka,Kc}$  & E$_{\rm up}$/k & A$_{\rm ij}$ &  Telescope & Beam & F$_{\rm eff}$ & B$_{\rm eff}$ & rms$^{(2)}$ & $\int \rm T_{\rm mb} dv$  & FWHM & Ref.$^{(3)}$ \\
& (GHz) & & (K)  & (s$ ^{-1}$) &  & size (\arcsec) & & & (mK) & (K.km s$^{-1}$) & (km s$^{-1}$) &\\
\hline
HDO & 481.7795 & 3$_{1,2}$-3$_{1,3}$ & 168 & 4.74 $\times$ 10$^{-5}$ &  HIFI 1a& 44.0 & 0.96 & 0.76 & 15 &$\le$ 0.12 & & a\\
HDO & 490.5966 & 2$_{0,2}$-1$_{1,1}$ & 66 & 5.25 $\times$ 10$^{-4}$ & HIFI 1a& 43.9 & 0.96 & 0.76 & 9 & 0.60 & 4.9 & a\\    
HDO & 509.2924 & 1$_{1,0}$-1$_{0,1}$ & 47 & 2.32 $\times$ 10$^{-3}$ & HIFI 1a& 42.3 & 0.96 & 0.76 &  9 & 1.08 & 5.8 &  a\\    
HDO & 599.9267 & 2$_{1,1}$-2$_{0,2}$ & 95 & 3.45 $\times$ 10$^{-3}$ & HIFI 1b & 35.9 & 0.96 & 0.75 &  9 & 0.83& 4.9 &  a\\    
HDO & 753.4112 & 3$_{1,2}$-3$_{0,3}$ & 168 & 5.90 $\times$ 10$^{-3}$ & HIFI 2b& 28.6 & 0.96 & 0.75 &  22 & 0.27 & 4.3 & a\\  
HDO & 848.9618 & 2$_{1,2}$-1$_{1,1}$ & 84 & 9.27 $\times$ 10$^{-4}$ & HIFI 3a & 25.4 & 0.96 & 0.75 &  20 & 0.54 & 5.3 & a\\  
HDO & 893.6387 & 1$_{1,1}$-0$_{0,0}$ & 43 & 8.35 $\times$ 10$^{-3}$ & HIFI 3b & 24.1 & 0.96 & 0.74 &  18 & 0.40 & 5.2 &  a\\  
HDO & 919.3109 & 2$_{0,2}$-1$_{0,1}$ & 66 & 1.56 $\times$ 10$^{-3}$ & HIFI 3b & 23.4 & 0.96 & 0.74 &  25 & 0.93 & 5.8 &  a\\  
HDO & 995.4115 & 3$_{0,3}$-2$_{1,2}$ & 131 & 7.04 $\times$ 10$^{-3}$ & HIFI 4a & 21.7 & 0.96 & 0.74 &  32 & 1.10 &  7.0 &  a\\ 
HDO & 1009.9447 & 2$_{1,1}$-1$_{1,0}$ & 95 & 1.56 $\times$ 10$^{-3}$ &  HIFI 4a& 21.0 & 0.96 & 0.74 & 29 & 0.38 & 6.5 & a\\
HDO & 1507.2610 & 3$_{1,2}$-2$_{1,1}$ & 168 & 6.58 $\times$ 10$^{-3}$ &  HIFI 6a& 14.1 & 0.96 & 0.72 & 257 &$\le$ 2.04 & &  a\\
HDO & 1625.4081 & 3$_{1,3}$-2$_{0,2}$ & 144 & 4.49 $\times$ 10$^{-2}$  &  HIFI 6b& 13.0 & 0.96 & 0.71 & 238 &$\le$ 1.89 & & a\\
\hline
HDO &  80.5783 & 1$_{1,0}$-1$_{1,1}$ & 47 & 1.32 $\times$ 10$^{-6}$ & IRAM-30m & 31.2 & 0.95 & 0.78 & 14 & 0.54 & 6.0  & b,d\\    
HDO & 225.8967 & 3$_{1,2}$-2$_{2,1}$ & 168 & 1.32 $\times$ 10$^{-5}$ & IRAM-30m & 11.1 & 0.91 & 0.54 & 34 & 2.15  & 7.4 &  b,d\\    
HDO & 241.5616 & 2$_{1,1}$-2$_{1,2}$ & 95 & 1.19 $\times$ 10$^{-5}$ & IRAM-30m & 10.4 & 0.91 & 0.51 & 23 & 2.27 & 6.8 & b\\      
HDO & 464.9245 & 1$_{0,1}$-0$_{0,0}$ & 22 & 1.69 $\times$ 10$^{-4}$ & JCMT & 10.8 & - & 0.5$^{(5)}$ & 63 & 5.50& 5.9 & c\\    
\hline
\hline
p-H$_2^{18}$O & 745.3202 & 2$_{1,1}$-2$_{0,2}$ & 136 & 6.83 $\times$ 10$^{-3}$ &  HIFI 2b & 28.4 & 0.96 & 0.75 & 18 & 0.34 & 5.2 &  a\\
p-H$_2^{18}$O & 970.2720 & 4$_{2,2}$-3$_{3,1}$ & 452 & 6.71 $\times$ 10$^{-4}$ &  HIFI 4a & 21.9 & 0.96 & 0.74 & 31 &$\le$ 0.23&  & a\\
p-H$_2^{18}$O & 994.6751 & 2$_{0,2}$-1$_{1,1}$ & 101 & 6.02 $\times$ 10$^{-3}$ &  HIFI 4a & 21.3 & 0.96 & 0.74 & 35 & 0.81 & 7.0 & a\\
p-H$_2^{18}$O & 1101.6983 & 1$_{1,1}$-0$_{0,0}$ & 53 & 1.79 $\times$ 10$^{-2}$ &  HIFI 4b & 19.2 & 0.96 & 0.74 & 44 & 0.63 & 5.3 &  a\\
p-H$_2^{18}$O & 1188.8631 & 4$_{2,2}$-4$_{1,3}$ & 452 & 2.73 $\times$ 10$^{-2}$ &  HIFI 5a& 17.8 & 0.96 & 0.64 & 125 &$\le$ 0.92 &  & a\\
p-H$_2^{18}$O & 1199.0056 & 2$_{2,0}$-2$_{1,1}$ & 194 & 1.76 $\times$ 10$^{-2}$ &  HIFI 5a& 17.7 & 0.96 & 0.64 & 90 &$\le$ 0.66 &  & a\\
p-H$_2^{18}$O & 1605.9625 & 4$_{1,3}$-4$_{0,4}$ & 395 & 3.71 $\times$ 10$^{-2}$ &  HIFI 6b& 13.2 & 0.96 & 0.71 & 339 &$\le$ 2.49 &   &a\\
\hline
o-H$_2^{18}$O & 489.0543 & 4$_{2,3}$-3$_{3,0}$ & 430$^{(4)}$  & 6.89 $\times$ 10$^{-5}$ &  HIFI 1a& 43.4 & 0.96 & 0.76 & 14 &$\le$ 0.10&   & a\\
o-H$_2^{18}$O & 547.6764 & 1$_{1,0}$-1$_{0,1}$ &   60$^{(4)}$  & 3.29 $\times$ 10$^{-3}$ &  HIFI 1a& 38.7 & 0.96 & 0.75 & 8 & 0.98 & 7.0  &a\\
o-H$_2^{18}$O & 1095.6274 & 3$_{1,2}$-3$_{0,3}$ & 249$^{(4)}$  & 1.62 $\times$ 10$^{-2}$ &  HIFI 4b& 19.4 & 0.96 & 0.74 & 49 & 0.38 & 3.0 & a\\
o-H$_2^{18}$O & 1136.7036 & 3$_{2,1}$-3$_{1,2}$ & 303$^{(4)}$  & 2.15 $\times$ 10$^{-2}$ &  HIFI 5a& 18.7 & 0.96 & 0.64 & 103 &$\le$ 0.76 &  & a\\
o-H$_2^{18}$O & 1181.3940 & 3$_{1,2}$-2$_{2,1}$ & 249$^{(4)}$ & 2.89 $\times$ 10$^{-3}$ &  HIFI 5a& 17.9 & 0.96 & 0.64 & 108 &$\le$ 0.79 &  & a\\
o-H$_2^{18}$O & 1633.4836 & 2$_{2,1}$-2$_{1,2}$ & 192$^{(4)}$ & 2.91 $\times$ 10$^{-2}$ &  HIFI 6b& 13.0 & 0.96 & 0.71 & 264 &$\le$ 1.94 &  & a\\
o-H$_2^{18}$O & 1655.8676 & 2$_{1,2}$-1$_{0,1}$ & 114$^{(4)}$ & 5.46 $\times$ 10$^{-2}$ &  HIFI 6b& 12.8 & 0.96 & 0.71 & 226 &$\le$ 1.66 &  & a\\
o-H$_2^{18}$O & 1719.2502 & 3$_{0,3}$-2$_{1,2}$ & 196$^{(4)}$ & 5.12 $\times$ 10$^{-2}$ &  HIFI 7a& 12.3 & 0.96 & 0.71 & 277 &$\le$ 2.04 &  & a\\
\hline
\hline
p-H$_2^{17}$O & 1107.1669 & 1$_{1,1}$-0$_{0,0}$ & 53 & 1.81 $\times$ 10$^{-2}$ &  HIFI 4b & 19.1  & 0.96 & 0.74 &  47 &  $\le$ 0.34 &  & a\\
\hline
o-H$_2^{17}$O & 552.0210 & 1$_{1,0}$-1$_{0,1}$ & 61$^{(4)}$  & 3.37 $\times$ 10$^{-3}$ &  HIFI 1a & 38.4 & 0.96 & 0.75 & 11 & 0.58 & 7.1  & a\\
\hline
\hline
HD$^{18}$O & 883.1894 & 1$_{1,1}$-0$_{0,0}$ & 42 & 7.96 $\times$ 10$^{-3}$ &  HIFI 3b & 24.0  & 0.96 & 0.75 &  22 &  $\le$ 0.16 &  & a\\
\hline
\end{tabular}
\end{center}
$^{(1)}$ The frequencies, the upper state energies (E$_{\rm up}$), and the Einstein coefficients (A$_{\rm ij}$) of HDO and H$_2^{18}$O come from the spectroscopic catalog JPL \citep{Pickett1998}. The ortho/para separation of H$_2^{18}$O was carried out in the CASSIS database.\\
$^{(2)}$ The rms is computed for a spectral resolution of 0.7\,km\,s$^{-1}$ for the HDO transitions and 0.6\,km\,s$^{-1}$ for the H$_2^{18}$O, H$_2^{17}$O, and HD$^{18}$O transitions.\\
$^{(3)}$ References : a) this work; b) \citet{Caux2011}; c) \citet{Stark2004}; d) \citet{Parise2005}\\
$^{(4)}$ The o-H$_2^{18}$O and o-H$_2^{17}$O upper energy levels quoted in the table assume that the fundamental 1$_{1,0}$-1$_{0,1}$ level lies at 34\,K and not at 0\,K.\\
$^{(5)}$ This value corresponds to the ratio between the beam efficiency and the forward efficiency.

\end{table*}%

\begin{figure}[!ht]
\includegraphics[scale=0.4]{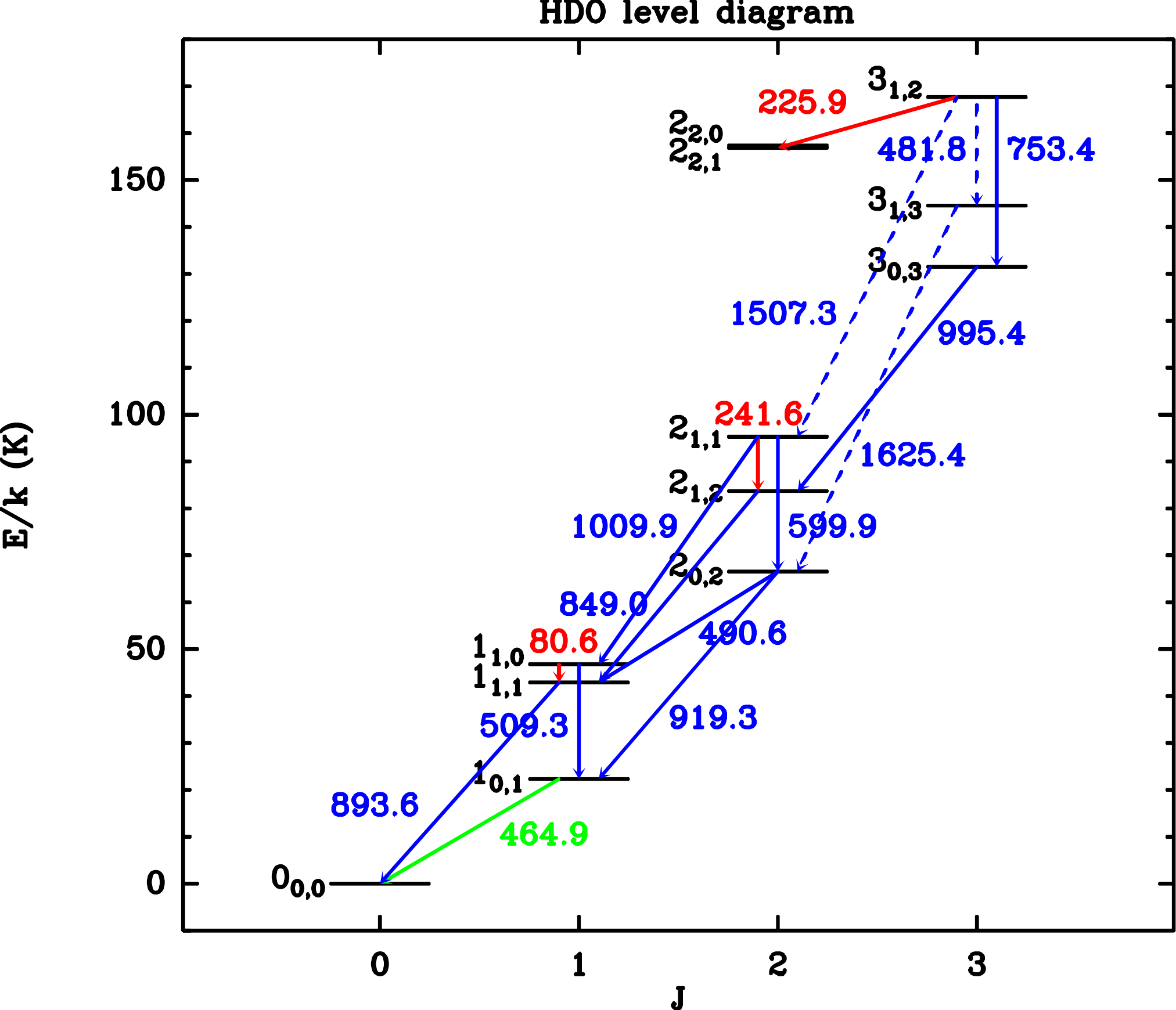}
\caption{Energy level diagram of the HDO lines. In \textit{red}, IRAM-30m observations; in \textit{green}, JCMT observation; in \textit{blue}, HIFI observations. The solid arrows show the detected transitions and the dashed arrows the undetected transitions. The frequencies are written in GHz.}
\label{leveldiagram}
\end{figure}

\

The antenna temperatures of the observations were converted to the $T\rm_{mb}$ scale, using the values of the main beam ($B_{\rm eff}$) and forward efficiencies ($F_{\rm eff}$) given in Table \ref{obs} with the usual relation:
\begin{equation}
T_{\rm mb} = T_{\rm A}^* \times \frac{F_{\rm eff}}{B_{\rm eff}}.
\end{equation}

Table \ref{obs} and the energy level diagram in Fig. \ref{leveldiagram} summarize the HDO, H$_2^{18}$O and H$_2^{17}$O transitions observed towards IRAS16293 with their flux or upper limit.
The 3$\sigma$ upper limits on the integrated line intensity are derived following the relation 
\begin{center}
\begin{equation}
 3 \sigma~ (\textrm{K.km~s$^{-1}$}) = 3 \times rms \times \sqrt{ 2 \times dv \times FWHM}
\end{equation}
\end{center}
with $rms$ (root mean square) in K, $dv$, the channel width, in km\,s$^{-1}$ and $FWHM$ (full width at half maximum) in km\,s$^{-1}$. We assume $FWHM = 5$\,km\,s$^{-1}$, which is the average emission linewidth.
The $FWHM$ given in Table \ref{obs} was determined with the CASSIS\footnote{CASSIS (http://cassis.cesr.fr) has been developed by IRAP-UPS/CNRS.} software by fitting the detected lines with a Gaussian.

Using the available spectroscopic databases JPL \citep{Pickett1998} and  CDMS \citep{Muller2001,Muller2005}, we carefully checked that none of our lines are contaminated by other species. 

\section{Modeling and results}

\subsection{Modeling}

The spherical Monte Carlo 1D radiative transfer code RATRAN \citep{Ratran}, which takes radiative pumping by continuum emission from dust into account, has been employed to compute the intensity of the molecular lines and the dust continuum.
An input model describing the molecular H$_2$ density, gas temperature, and velocity field profiles is required, in order to define the spherical region into many radial cells. The code applies a Monte Carlo method to iteratively converge on the mean radiation field, J$_{\nu}$. Level populations of HDO, H$_2^{18}$O, H$_2^{17}$O, and HD$^{18}$O can then be calculated, once J$_{\nu}$ is determined for every cell. These level populations are required to map the emission distribution throughout the cloud. 

\begin{figure}[!t]
\begin{center}
\includegraphics[scale=0.3]{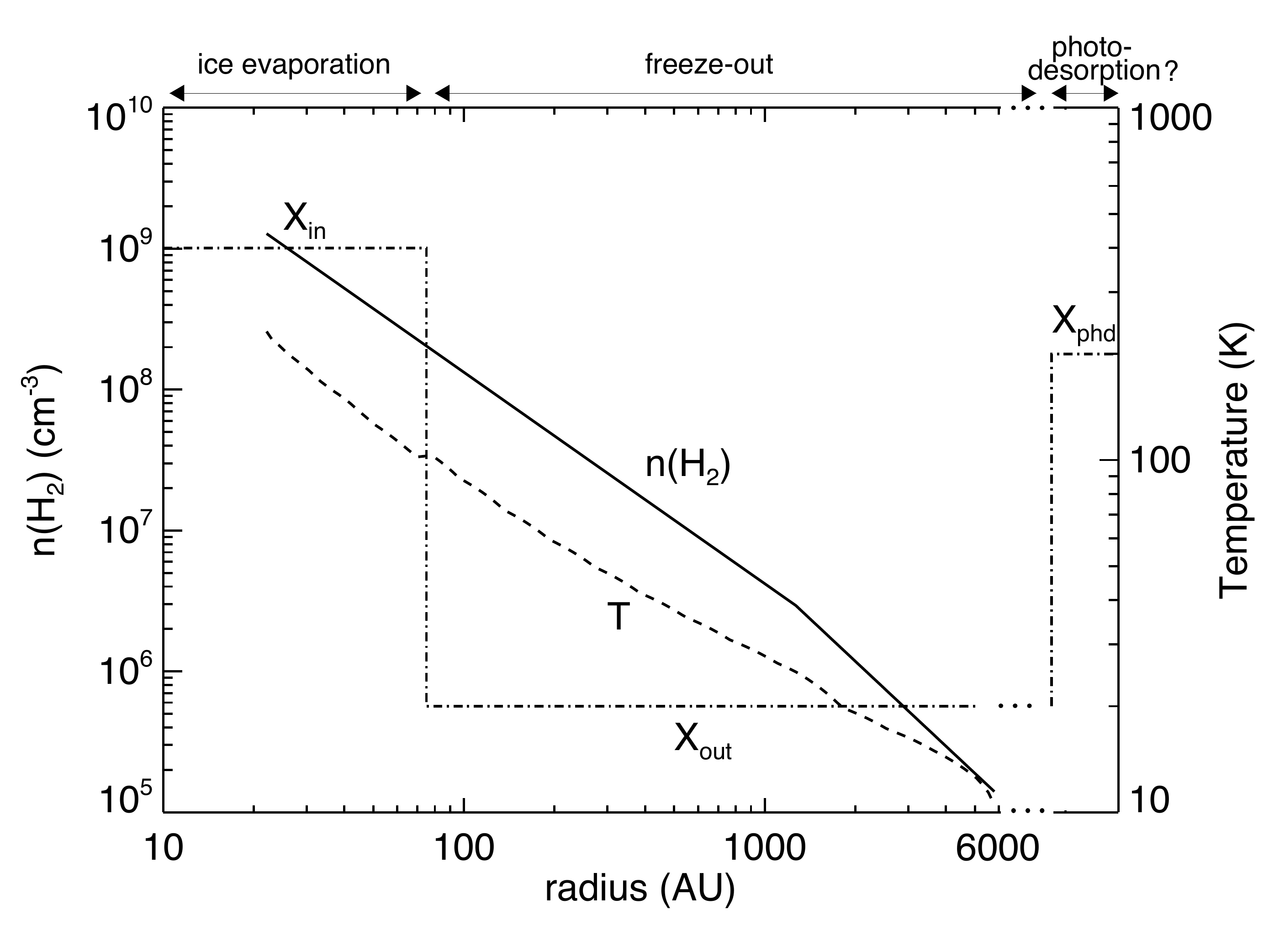}
\caption{H$_2$ density (solid line) and gas temperature (dashed line) structure of IRAS16293 determined by \citet{Crimier2010}.
The expected abundance profile of water is added (dashed-dotted line) on an arbitrary Y-scale. In the colder envelope ($T$$<$100\,K), the water molecules are trapped on the grain mantles, whereas in the inner part of the envelope, they are released in gas phase by thermal heating, leading to an enhancement of the abundance. In the outer part of the molecular cloud (A$_V$ $\sim$ 1-4), the water abundance can also increase by photodesorption mechanisms. A freeze-out timescale much longer than the protostellar lifetime can also lead to an enhanced abundance when the density is low ($\sim$ 10$^4$ cm$^{-3}$).}
\label{structure}
\end{center}
\end{figure}

The source structure used in the modeling was determined by \citet{Crimier2010} with a radius extending from 22 AU to 6100 AU (see Fig. \ref{structure}); however, the structure in the hot corino region (T $\geq$ 100 K) may be uncertain since disks probably exist in the inner part of the protostellar envelopes. Nevertheless, since the disk characteristics are unknown, we keep the structure as it is. 
The radial velocity $ v_r = \sqrt{ 2 G M / r}$ (where $M$ is the stellar mass, $G$ the gravitational constant and $r$ the radius) is calculated for a stellar mass of 1 $M_\odot$. 
For a higher mass ($\sim$ 2 $M_\odot$), the line widths become too broad to reproduce the line profiles. Moreover, the value of 1 $M_\odot$ agrees with the mass of the core A, while the mass of the core B is at most about 0.1 $M_\odot$ \citep{Ceccarelli2000,Bottinelli2004,Caux2011}.
For a radius greater than 1280 AU, the envelope is considered as static (the velocity is fixed at 0). This radius corresponds to a change in the slope of the density profile, marking the transition of the collapsing/static envelope \citep{Shu1977}, and its value has been determined by \citet{Crimier2010}.
To reproduce the widths of the absorption lines, the turbulence Doppler b-parameter (equal to 0.6 $\times$ $FWHM$) is fixed at 0.3\,km\,s$^{-1}$. If a lower (respectively higher) value is adopted, the modeled absorption lines become too narrow (respectively too broad) compared with the observations.
As source A is more massive than source B, we considered that the structure is centered on core A. 
According to interferometric data of the HDO 3$_{1,2}$-2$_{2,1}$ line at 226 GHz, deuterated water is mainly emitted by core A \citep{Jorgensen2011}. 
Therefore, the assumption of a 1D modeling centered on source A seems quite reasonable. Because the IRAM observations were pointed on the IRAS16293 B source and not on core A and the beam is quite small ($\sim$ 11$\arcsec$) at 226 and 242 GHz, we carefully convolved  the resulting map with the telescope beam profile centered on core B to get the model spectra.

To fit the continuum observed with HIFI from band 1 to band 4, the dust opacity as function of the frequency has to be constrained by a power-law emissivity model  :
\begin{equation}
\kappa = \kappa_0 \left( \frac{\nu}{\nu_0}\right)^{\beta}
\end{equation}
with $\beta=1.8$, $\kappa_0 =15$ cm$^2$/g$_{\rm dust}$, and $\nu_0 = 10^{12}$ Hz, which has been used as input in the RATRAN modeling.
Using this emissivity model and the source structure described above, the continuum of the HDO 1$_{0,1}$-0$_{0,0}$ fundamental line at 465 GHz is predicted at $\sim$1 K. The continuum determination at this frequency is essential for the modeling because this line shows a deep and narrow self-absorption.
The modeled continuum is shown in Figs. \ref{comp_abs} and \ref{ratran_outercell} for the two HDO absorption lines, whereas it has been subtracted for all the lines that only present emission.
The profiles obtained are resampled at the spectral resolution of the observations. Smoothing was applied on some observations when the line is undetected or weakly detected (see Figs. \ref{ratran_outercell} and  \ref{fig_h218o}). For the determination of the best fit parameters by $\chi^2$ minimization (see Sect. 3.3), the spectra are also resampled to a same spectral resolution for all the lines (0.7\,km\,s$^{-1}$  for HDO and 0.6\,km\,s$^{-1}$ for H$_2^{18}$O, H$_2^{17}$O, and HD$^{18}$O, which correspond to the lowest spectral resolution of the HIFI data).

Previous studies \citep[e.g.][]{vanKempen2008} have shown that in protostellar envelopes, the dust can become optically thick, preventing water emission from deep in the envelope to escape. The dust continuum could therefore hide the higher frequencies HDO and H$_2^{18}$O lines coming from the hot corino, depending on the source structure and the dust emissivity model chosen.
In fact, we notice here that the dust optical depth is always lower than 1 for all the frequencies lower than 1500 GHz.
For the transitions quoted in Table \ref{obs} with frequencies higher than 1500 GHz, the dust opacity only exceeds 1 for temperatures well over 100 K.
Consequently, in the case of IRAS16293, the dust opacity should not hide the higher frequencies HDO and  H$_2^{18}$O lines as the dust opacity only becomes thick in the very deep part of the inner envelope.

\subsection{HDO collisional rate coefficients}

The HDO collisional rates used in this study have recently been computed by \citet{Faure2011} for para-H$_2$($J_2=0, 2$) and ortho-H$_2$ ($J_2=1$) in the temperature range 5$-$300~K, and for all the transitions with an upper energy level less than 444~K. The methodology used by Faure et al. (2011) is described in detail in \citet{Scribano2010} and \citet{Wiesenfeld2011}. These authors present detailed comparisons between the three water-hydrogen isotopologues H$_2$O-H$_2$, HDO-H$_2$, and D$_2$O-H$_2$. Significant differences were observed in the cross sections and rates and were attributed to symmetry, kinematics, and intramolecular (monomer) geometry effects. Moreover, in the case of HDO, rate coefficients with H$_2$  were found to be significantly larger, by up to three orders of magnitude, than the (scaled) H$_2$O-He rate coefficients of \citet{Green1989}, which are currently employed in astronomical models \citep[see Fig.~2 of][]{Faure2011}. A significant impact of the new HDO rate coefficients is thus expected in the determination of interstellar HDO abundances, as examined in Sect. 3.3. In the following, we assume that the ortho-to-para ratio of H$_2$ is at local thermodynamic equilibrium (LTE) in each cell of the envelope. The collisional rates for para-H$_2$ were also summed and averaged by assuming a thermal distribution of $J_2=0, 2$.

\subsection{Determination of the HDO abundance}

In a first step, we ran a grid of models with one abundance jump at 100\,K, in the so-called hot corino \citep{Ceccarelli1996,Ceccarelli2000b}.
The abundance of water is expected to be higher in the hot corino than in the colder envelope, since the water molecules contained in the icy grain mantles are released in gas phase, when the temperature is higher than the sublimation temperature of water, $\sim$100\,K \citep{Fraser2001}. Thereafter, the inner abundance ($T$ $\ge$ 100\,K) will be designated by $X_{\rm in}$ and the outer abundance  ($T$ $<$ 100\,K) by $X_{\rm out}$. Both are free parameters and their best fit values are determined by a $\chi^2$ minimization.
To take the line profile into account, the $\chi^2$ is computed from the observed and modeled spectra resampled at a same spectral resolution (0.7\,km\,s$^{-1}$ for HDO and 0.6\,km\,s$^{-1}$ for H$_2^{18}$O) according to the following formalism:
\begin{equation}
\centering
 \chi^2 = \sum_{ \rm i=1}^{ N} \sum_{\rm j=1}^{ n_{\rm chan}} \frac{(T_{\rm obs,ij}-T_{\rm mod,ij})^2}{ rms_{\rm i}^2 + (Cal_{\rm i} \times T_{\rm obs,ij})^2}
 \label{chi2_eq}
\end{equation}
with $N$ the number of lines i, $n_{\rm chan}$ the number of channels j for each line, $T_{\rm obs,ij}$ and $  T_{\rm mod,ij}$ the intensity observed and predicted by the model respectively in channel j of the line i, $ rms_{ \rm i}$ the rms at the 0.7\,km\,s$^{-1}$ (or 0.6\,km\,s$^{-1}$) spectral resolution (given in Table 1), and $Cal_{ \rm i}$ the calibration uncertainty. We assumed an overall calibration uncertainty of 15\% for each detected line. 

\begin{figure}[!ht]
\begin{center}
\includegraphics[scale=0.33]{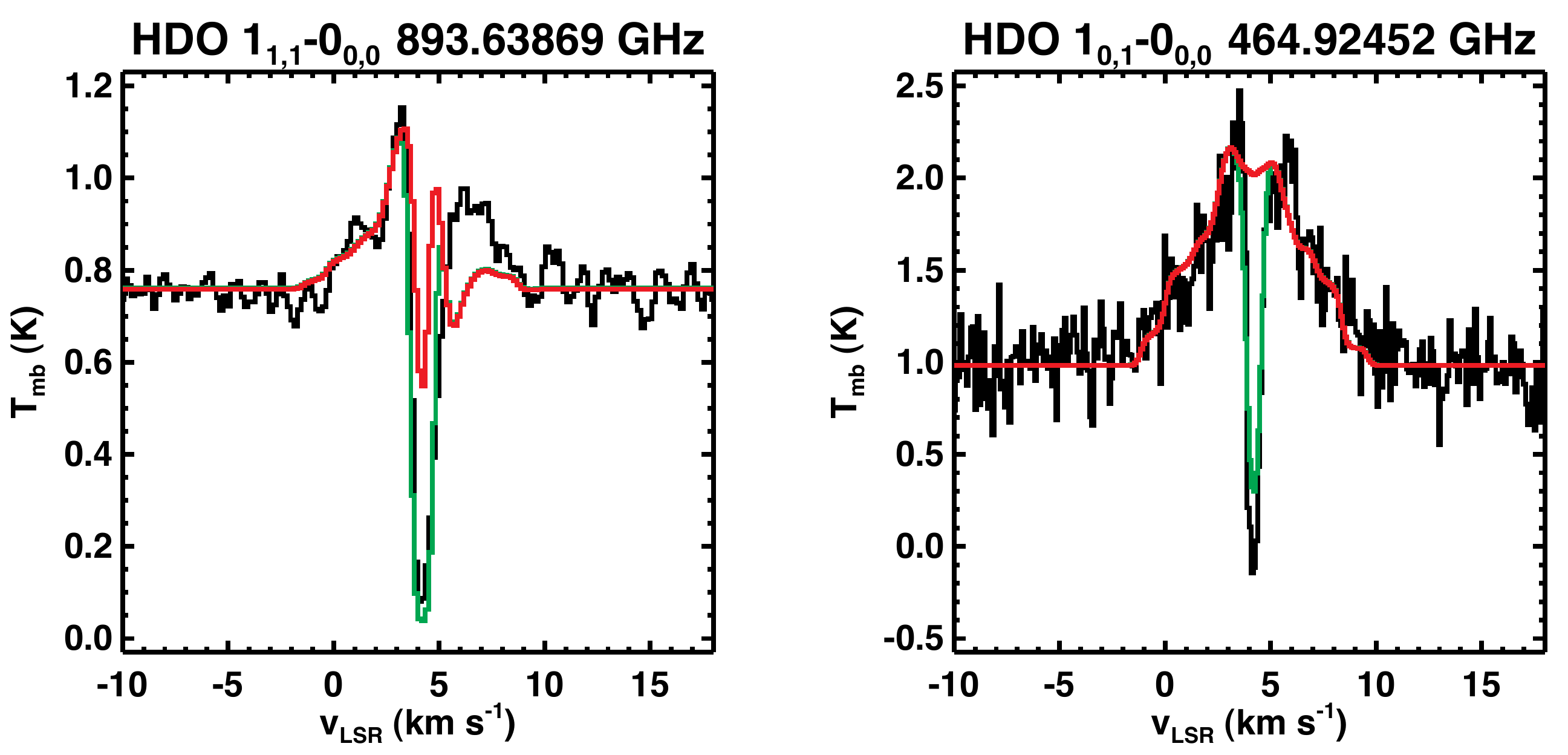}
\caption{In black: HDO 1$_{1,1}$-0$_{0,0}$ and 1$_{1,0}$-0$_{0,0}$ absorption lines observed at 894 GHz with HIFI  and at 465 GHz with JCMT, respectively.
In red: HDO modeling without adding the absorbing layer. 
In green: HDO modeling when adding an absorbing layer with a HDO column density of $\sim$ 2.3 $\times$ 10$^{13}$ cm$^{-2}$.
The continuum for both the 894 GHz and 465 GHz lines refers to SSB data.}
\label{comp_abs}
\end{center}
\end{figure}

The best fit result for this grid of models gives an inner abundance $X_{\rm in}$ = 1.9 $\times$ 10$^{-7}$ and an outer abundance $X_{\rm out}$ = 5 $\times$ 10$^{-11}$ with a reduced $\rm\chi^2$ of 3.2.
However this model (as well as the other models of this grid) predicts absorption lines that are weaker than observed (see Fig. \ref{comp_abs}). To correctly reproduce the depth of the absorption in the 465 and 894 GHz lines, while not introducing extra emission in the remaining ones, it is necessary to add an absorbing layer in front of the IRAS16293 envelope. We notice that the results mainly depend on the HDO column density of the layer, and they are insensitive to the density (for densities lower than $\sim 10^5$ cm$^{-3}$) and to the temperature (for temperatures lower than $\sim$~30\,K).
The HDO column density of the layer must be about 2.3 $\times$ 10$^{13}$ cm$^{-2}$.
To understand from where the HDO absorptions arise, we can consider two different options.
\begin{itemize}
\item 
First, the source structure should be extended to a higher outer radius to include the whole molecular cloud.
If the absorbing layer comes from the surrounding gas of the molecular cloud harboring IRAS16293 and if we assume that the HDO abundance remains constant as in the outer envelope, this would imply a surrounding H$_2$ column density of about 2.9 $\times$ 10$^{23}$ cm$^{-2}$. But this value is considerably too high for a molecular cloud. 
The N(H$_2$) column density of the $\rho$ Oph cloud has been estimated at $\sim$ 1.5 $\times$ 10$^{22}$ cm$^{-2}$ by \citet{VanDishoeck1995} and at $\gtrsim$ 5 $\times$ 10$^{22}$ cm$^{-2}$ by \citet{Caux1999}. Adding a molecular cloud with these characteristics to the structure determined by \citet{Crimier2010} does not allow deep absorption lines to be modeled.
This hypothesis is therefore insufficient to explain the deep HDO absorption lines. 
\item
Consequently, the only way to explain the absorption lines consists in assuming a drop abundance structure. Such a structure in the low-mass protostars has already been inferred for several molecules like CO and H$_2$CO \citep{Schoier2004,Jorgensen2004,Jorgensen2005}. 
The raising of the abundance in the outermost regions of the envelope is explained by a longer depletion timescale than the lifetime of the protostars ($\sim$10$^{4}$-10$^5$ years) for H$_2$ densities lower than $\sim$10$^4$-10$^5$ cm$^{-3}$ \citep[e.g.,][]{Caselli1999,Jorgensen2004}.
On the other hand, \citet{Hollenbach2009} have shown that, at the edge of molecular clouds, the icy mantles are photodesorbed by the UV photons, giving rise to an extended layer with a higher water abundance for a visual extinction $A_{\rm V}$$\sim$ 1-4 mag (for $G_0=1$). 
Assuming an $A_{\rm V}$ of $\sim$1-4 mag and the relation $N(\textrm{H$_2$})/A_{\rm V}$ = 9.4 $\times$ 10$^{20}$ cm$^{-2}$ mag$^{-1}$ \citep{Frerkin1982}, the abundance of HDO in this water-rich layer created by the photodesorption of the ices should be about 6~$\times$~10$^{-9}$~-~2.4~$\times$~10$^{-8}$.  
We see thereafter, when analyzing the H$_2^{18}$O lines, that this hypothesis is nicely consistent with the theoretical predictions by \citet{Hollenbach2009}.
For water, both effects may play a role.
\end{itemize}

\begin{figure*}[!ht]
\begin{center}
\includegraphics[scale=0.65]{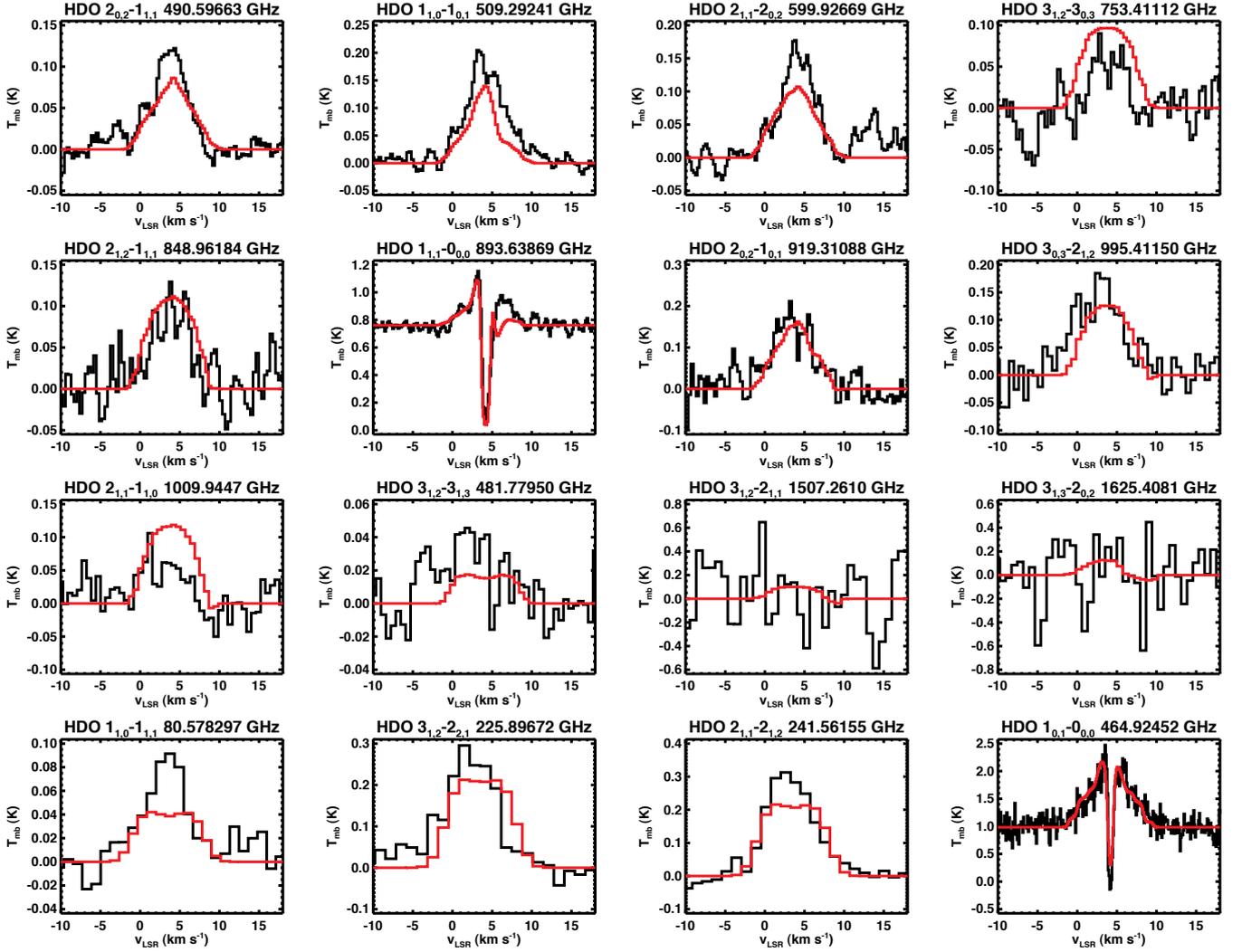}
\caption{In black: HDO lines observed with HIFI, IRAM, and JCMT.
In red: best-fit model obtained when adding an absorbing layer with an HDO column density of $\sim$ 2.3 $\times$ 10$^{13}$ cm$^{-2}$ to the structure (see details in text).
The best-fit inner abundance is 1.7 $\times$ 10$^{-7}$ and the best-fit outer abundance is 8 $\times$ 10$^{-11}$.
The continuum shown for both the 894 GHz  and 465 GHz lines, refers to SSB data.}
\label{ratran_outercell}
\end{center}
\end{figure*}

\begin{figure}[!ht]
\begin{center}
\includegraphics[scale=0.3]{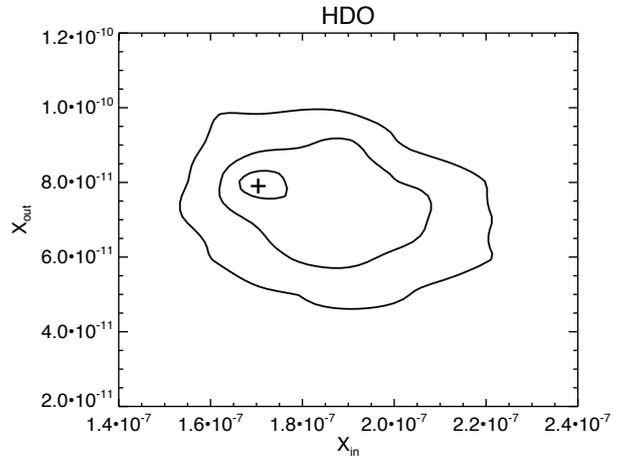}
\caption{$\rm\chi^2$ contours at 1$\rm \sigma$, 2$\rm \sigma$, and 3$\rm \sigma$ obtained when adding an absorbing layer with a HDO column density of 2.3 $\times$ 10$^{13}$ cm$^{-2}$ to the structure. The results are obtained with the collisional coefficients with ortho and para-H$_2$ determined by \citet{Faure2011}. The best-fit model is represented by the symbol ``+''. The reduced  $\chi^2$ is about 2.4.}
\label{chi2}
\end{center}
\end{figure}

As the absorption of the two fundamental lines is reproduced with this outer component (see Fig. \ref{comp_abs}), we ran a second grid of models adding this absorbing layer.
The best fit is obtained for $X_{\rm in}$ = 1.7 $\times$ 10$^{-7}$ and $X_{\rm out}$ = 8 $\times$ 10$^{-11}$ with a $\rm\chi_{red}^2$ of 2.4.
With the current parameters, a simultaneous fit of all the HIFI, IRAM, and JCMT data is presented in Fig. \ref{ratran_outercell}.
The contours of $\chi^2$ at 1$\rm \sigma$, 2$\rm \sigma$, and 3$\rm \sigma$, which respectively represent a confidence of 68.26\%, 95.44\%, and 99.73\% of enclosing the true values of $X_{\rm in}$ and $X_{\rm out}$, are shown in Fig. \ref{chi2}, computed with the method of \citet{Lampton1976}. 
The contours at 1$\rm \sigma$, 2$\rm \sigma$, and 3$\rm \sigma$ correspond respectively to $\chi^2$=$\rm \chi^2_{min}$+2.3, $\chi^2$=$\rm \chi^2_{min}$+6.17, and $\chi^2$=$\rm \chi^2_{min}$+11.8 when the number of adjustable parameters is two (here $X_{\rm in}$ and $X_{\rm out}$). The confidence intervals of $X_{\rm in}$ and $X_{\rm out}$ at 3$\sigma$ are 1.53 $\times$ 10$^{-7}$ -  2.21 $\times$ 10$^{-7}$  and 4.6 $\times$ 10$^{-11}$ - 1.0 $\times$ 10$^{-10}$, respectively.

\begin{figure}[!ht]
\begin{center}
\includegraphics[scale=0.3]{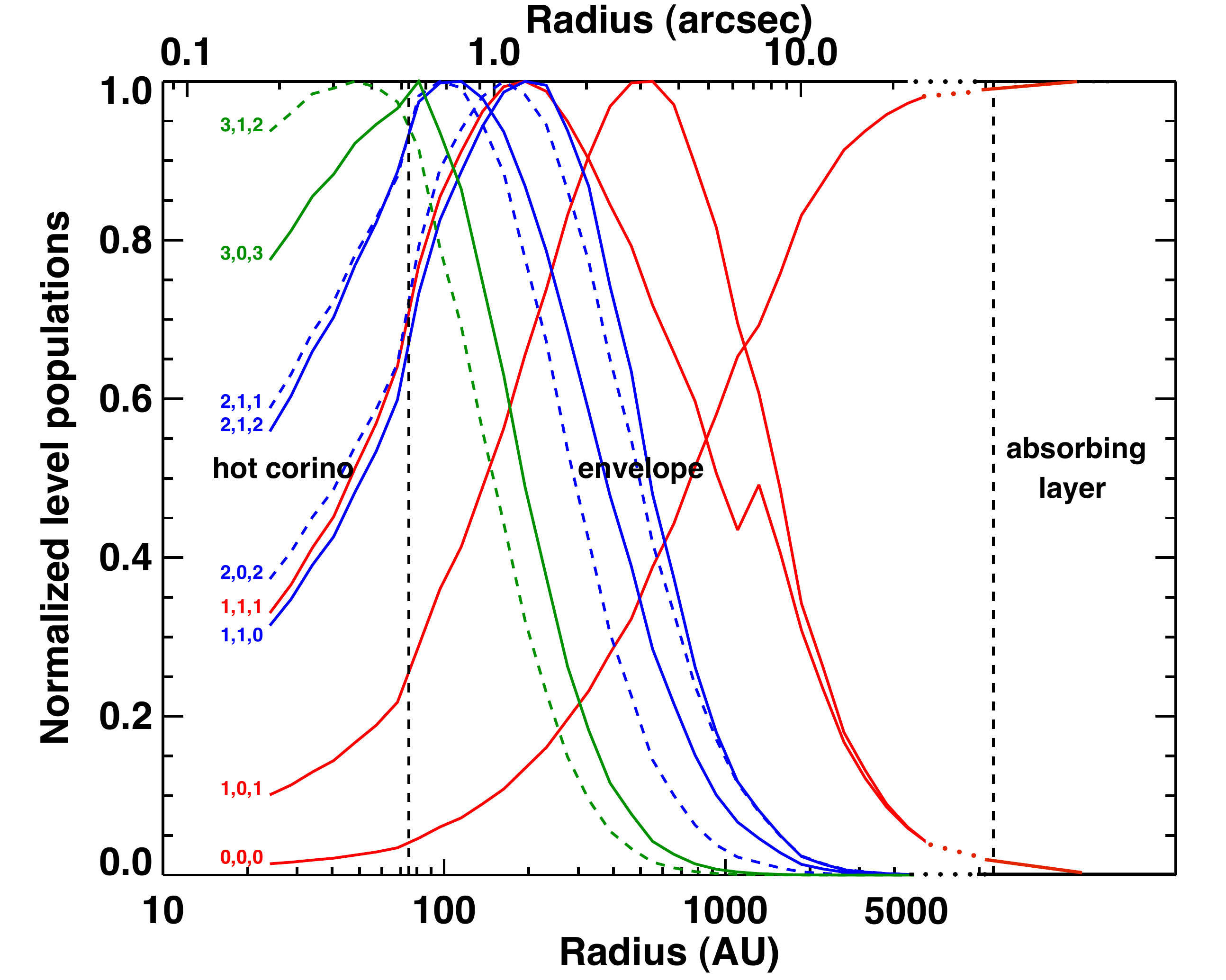}
\caption{Normalized level populations of HDO computed by RATRAN, as a function of the radius of the protostar envelope. To avoid confusion, the population of levels 2$_{0,2}$, 2$_{1,1}$, and 3$_{1,2}$ are indicated by dashed lines. 
The HDO 1$_{1,1}$ level population jump visible at $\sim$1300 AU is an artifact due to the change of the velocity field at the interface between the infalling envelope and the static envelope.
}
\label{pop}
\end{center}
\end{figure}

In the hot corino, most of the HDO lines are optically thick, except the 81, 226, 242, and 482 GHz transitions that show an opacity $\lesssim$ 1.  Since the emission from most of the HDO lines originates both in the hot corino and the outer envelope, the use of a simple rotational-diagram analysis to estimate the total column density of HDO is not appropriate.
Figure \ref{pop} represents the normalized populations of some levels (up to 3$_{1,2}$) as a function of the distance from the center of the protostar, computed by RATRAN through the equation of statistical equilibrium. These level populations are then used to determine the emission distribution of deuterated water by using a ray-tracing method. This figure also shows that the absorbing layer is mainly constrained by the ground state transitions as 1$_{1,1}$-0$_{0,0}$, and 1$_{0,1}$-0$_{0,0}$.

The 81 GHz 1$_{1,0}$-1$_{1,1}$ transition observed at IRAM-30m is not reproduced well by the best-fit model.  
The disagreement could be due to a blending effect at this frequency. Nevertheless, it is ruled out for the species included in the JPL and CDMS databases. 
Calibration could also be questioned at this frequency. But it seems rejected since the profile obtained with the best-fit model differs from the observing profile. 
Another explanation could come from the assumption of spherical symmetry. Interferometric data of the HDO 3$_{1,2}$-2$_{2,1}$ transition show that, in addition to the main emission coming from the source A, a weaker component of HDO is emitted at about 6\arcsec~from core~A (see Fig. 20 in \citet{Jorgensen2011}). The beam size of the IRAM-30m telescope at 81 GHz encompasses this component and could explain the lack of predicted flux as well as the different line profile. This hypothesis seems consistent with the lower frequency HIFI lines (491, 509, and 600 GHz) for which the best-fit model misses flux.

The derived abundances for HDO agrees with the 3$\sigma$ upper limit (0.16\,K\,km\,s$^{-1}$) of the HD$^{18}$O 1$_{1,1}$-0$_{0,0}$ transition observed in our spectra at 883 GHz. Assuming an HD$^{16}$O/HD$^{18}$O ratio of 500 \citep{WIlson1994}, value at the galactocentric distance of IRAS16293, the modeling predicts an integrated intensity of 0.15\,K\,km\,s$^{-1}$.

To obtain consistent profiles of the model compared with the observations, the LSR velocity, $V_{\rm LSR}$, is about 4.2\,km\,s$^{-1}$ for the HIFI and JCMT lines, whereas it is about 3.6\,km\,s$^{-1}$ for the three IRAM lines. This discrepancy of the velocity in the modeling could be explained by the origin of the emission of the lines.
The 4.2\,km\,s$^{-1}$ velocity component could represent the velocity of the envelope. For example, the velocity of the absorption lines of the fundamental transitions of D$_2$O tracing the cold envelope of IRAS16293 and detected by JCMT \citep{Butner2007} and by HIFI \citep{Vastel2010} is 4.15 and 4.33\,km\,s$^{-1}$, respectively. Even if some HIFI lines trace both the inner part and the outer part of the envelope, the velocity of the outer envelope should dominate because of a smaller line widening in the outer envelope.
In contrast, the flux of IRAM lines at 226 and 242 GHz only trace the hot corino (see Figs. \ref{leveldiagram} and \ref{pop}). Their velocity ($\sim$ 3.6\,km\,s$^{-1}$) is lower, in agreement with the velocity of the core A ($V_{\rm LSR}$ $\sim$ 3.9\,km\,s$^{-1}$).

As previous studies regarding HDO used the collisional coefficients computed by \citet{Green1989}, we have run a grid of models with these rates for a comparison.
The simultaneous best-fit of all the transitions using these rates is obtained for an inner abundance $X_{\rm in}$ = 2.0 $\times$ 10$^{-7}$ and an outer abundance $X_{\rm out}$ = 1 $\times$ 10$^{-11}$. The inner abundance is similar to what we found above with the collisional coefficients determined by \citet{Faure2011}. However, the outer abundance is a factor of 8 lower. The $\rm\chi^2$ also gives a higher value ($\rm \chi^2_{red}$ = 2.8), therefore showing that the observations are better reproduced with rates with H$_2$ than with He.

\subsection{Determination of the water abundance}

\begin{figure}[!t]
\begin{center}
\includegraphics[scale=0.3]{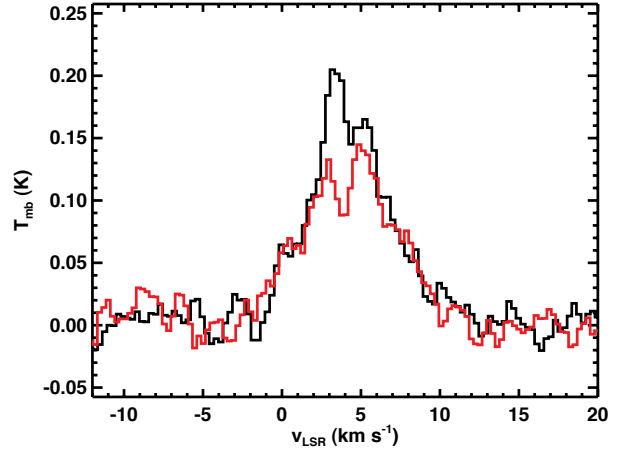}
\caption{Comparison of the profiles of the HDO (in black) and H$_2^{18}$O (in red) 1$_{1,0}$-1$_{0,1}$ transitions at 509 and 548 GHz, respectively.}
\label{width}
\end{center}
\end{figure}

To determine the H$_2$O abundance throughout the envelope, we used all the H$_2^{18}$O transitions in the HIFI range (see Table \ref{obs}). The only transition observable in the TIMASSS spectral survey is contaminated by the CH$_3$OCH$_3$ species at 203.4 GHz. The profiles of the observed H$_2^{16}$O lines suggest, by the presence of wings, that the lines are contaminated by the outflows. Unlike the H$_2^{16}$O lines that show a width $\gtrsim$ 10\,km\,s$^{-1}$, the H$_2^{18}$O line widths are similar to the HDO lines (see Fig. \ref{width}, $\sim$ 5 km\,s$^{-1}$) and no wings are seen in the H$_2^{18}$O detected transitions (see Fig. \ref{fig_h218o}). Consequently, the outflow does not probably contribute to a large extent to the emission of the H$_2^{18}$O lines. 

\begin{figure}[!t]
\begin{center}
\includegraphics[scale=0.3]{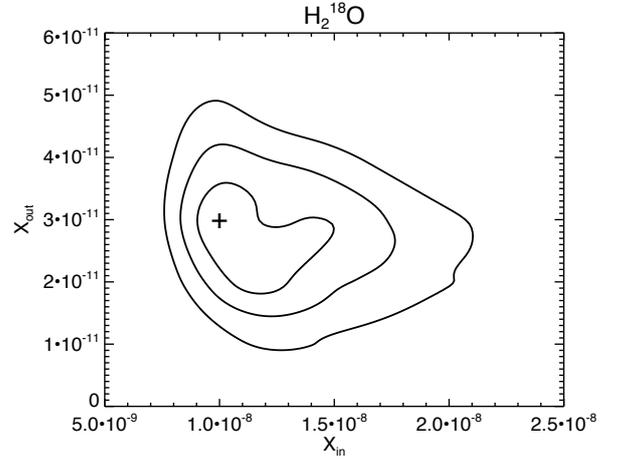}
\caption{$\rm\chi^2$ contours at 1$\rm \sigma$, 2$\rm \sigma$, and 3$\rm \sigma$ obtained for H$_2^{18}$O. The best-fit model is represented by the symbol ``+''. The reduced $\chi^2$ is about 1.9.}
\label{chi2_h218o}
\end{center}
\end{figure}

\begin{figure}[!ht]
\begin{center}
\includegraphics[scale=0.3]{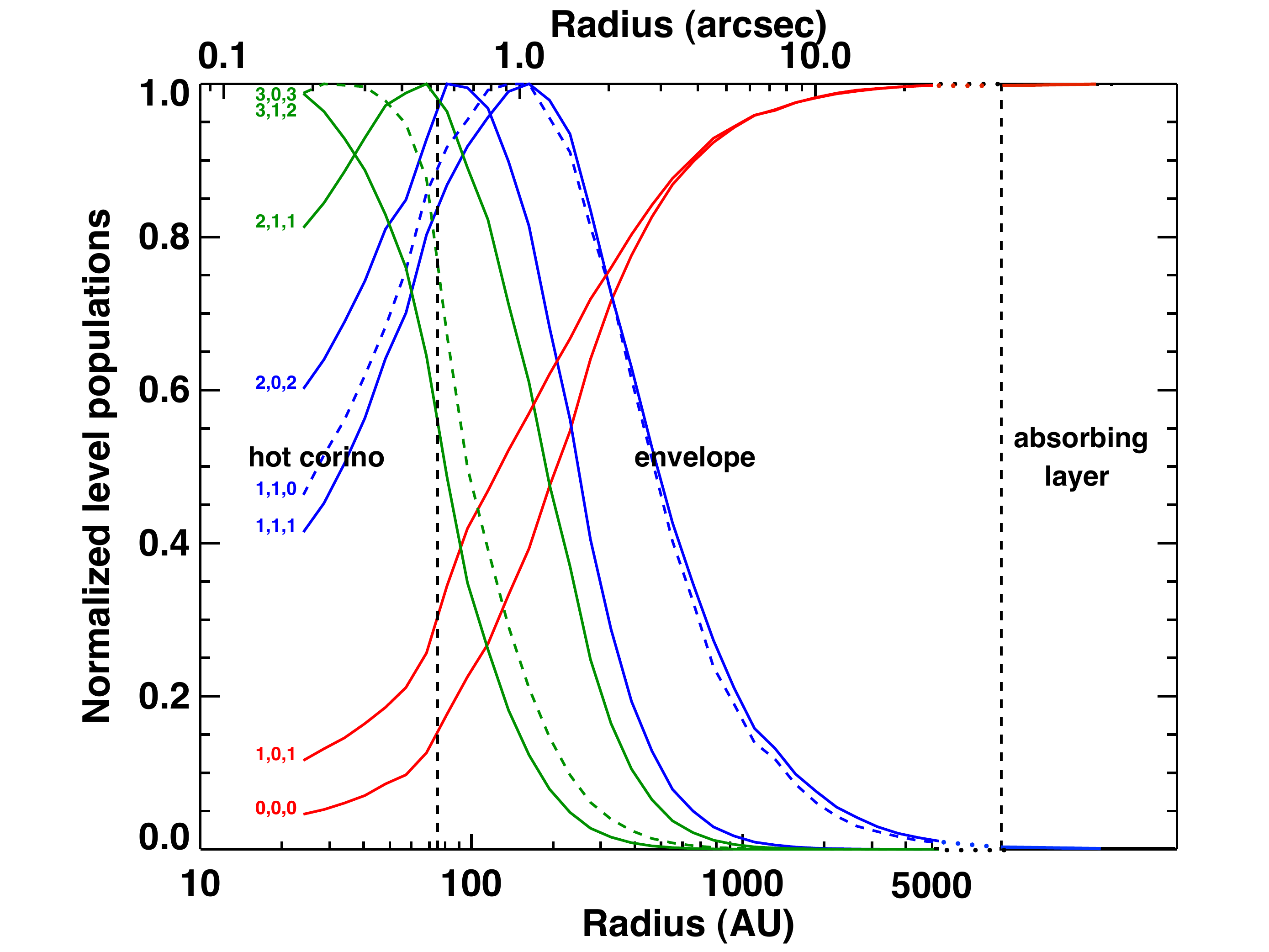}
\caption{Normalized level populations of H$_2^{18}$O computed by RATRAN, as a function of the radius of the protostar envelope. To avoid confusion, the population of levels 1$_{1,0}$ and 3$_{0,3}$ are indicated by dashed lines. }
\label{pop_h218o}
\end{center}
\end{figure}

\begin{figure*}[!ht]
\begin{center}
\includegraphics[scale=0.65]{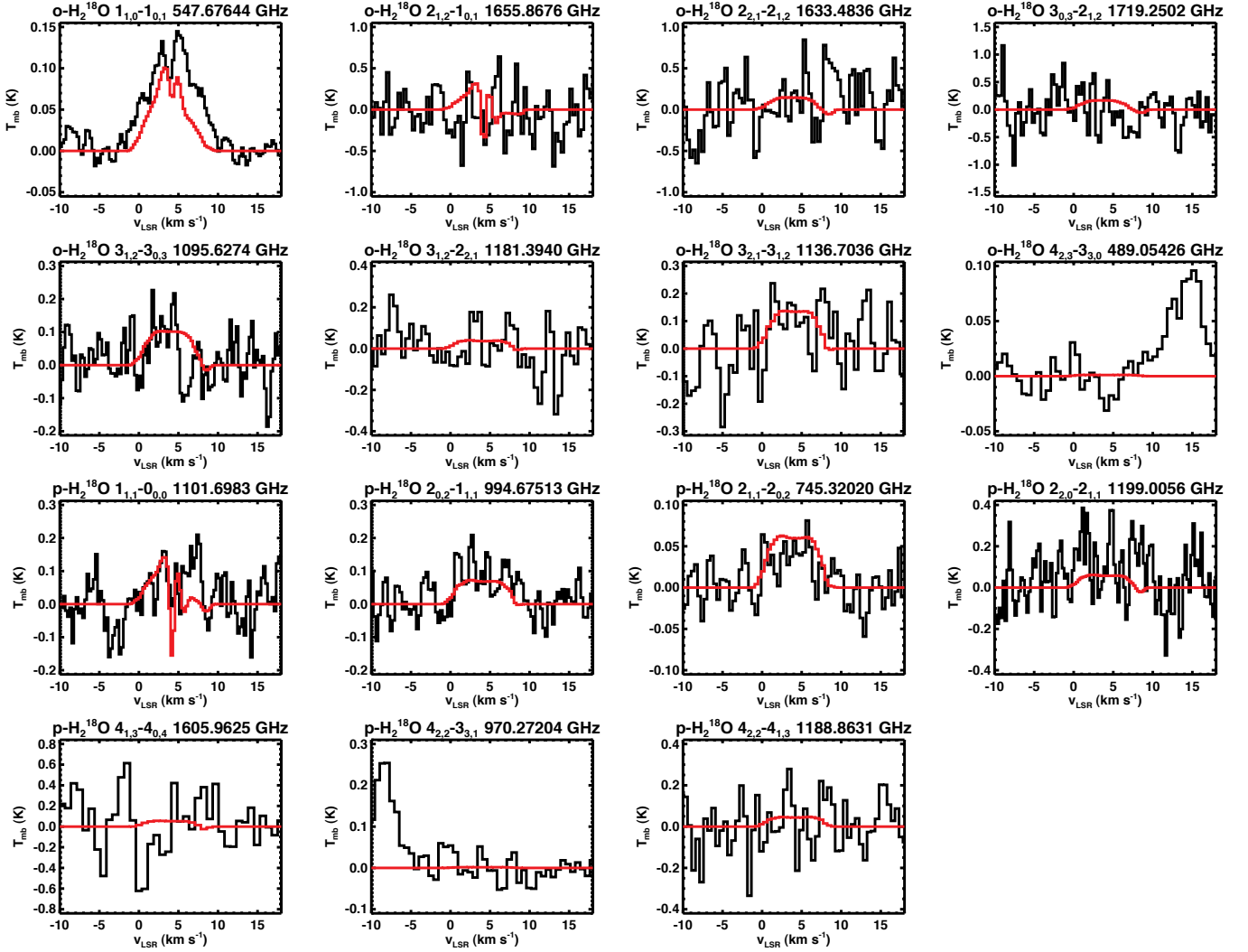}
\caption{In black: H$_2^{18}$O lines observed with HIFI.
In red: best-fit model obtained when adding an absorbing layer with a H$_2^{18}$O column density of 1 $\times$ 10$^{12}$ cm$^{-2}$.
The inner abundance is 1 $\times$ 10$^{-8}$ and the outer abundance is 3 $\times$ 10$^{-11}$. The detected line at a velocity of 15\,km\,s$^{-1}$ in the panel of the 489 GHz transition is CH$_3$OH at 489.0368 GHz. Also, the line observed at a velocity of -10\,km\,s$^{-1}$ in the panel of the 970 GHz transition is p-H$_2$O at 970.3152 GHz.}
\label{fig_h218o}
\end{center}
\end{figure*}

We used the H$_2$O collisional coefficients determined by \citet{Faure2007}
and assumed a H$_2^{16}$O/H$_2^{18}$O ratio equal to 500 \citep{WIlson1994}, as well as a H$_2^{18}$O  ortho/para ratio of 3.
As for HDO, in a first step, we ran a grid of models for different inner and outer abundances without adding the absorbing layer. 
The best-fit parameters of the $\rm \chi^2$ minimization are $X_{\rm in}$(H$_2^{18}$O) = 1 $\times$ 10$^{-8}$ and $X_{\rm out}$(H$_2^{18}$O) = 3 $\times$ 10$^{-11}$ and the reduced $\rm \chi^2$ is about 1.9.
The inner abundance of water is therefore 5 $\times$ 10$^{-6}$, whereas the outer abundance is 1.5 $\times$ 10$^{-8}$.
Figure \ref{chi2_h218o} shows the $\chi^2$ contours at 1$\rm \sigma$, 2$\rm \sigma$, and 3$\rm \sigma$. A calibration uncertainty of 15\% has been assumed for the 548 GHz transition and the tentatively detected ones (1096, 1102, 994, and 745 GHz). At 3$\rm \sigma$, the outer abundance of H$_2^{18}$O varies between 9 $\times$ 10$^{-12}$ and 4.9 $\times$ 10$^{-11}$, whereas the inner abundance is in the interval 7.6 $\times$ 10$^{-9}$ - 2.1 $\times$ 10$^{-8}$.
Using the best-fit values found for $X_{\rm in}$ and $X_{\rm out}$,  it is necessary, in order to reproduce the weak absorption of the 1$_{1,0}$-1$_{0,1}$  line at 548 GHz, to add an absorbing layer with a H$_2^{18}$O column density of about 1 $\times$ 10$^{12}$ cm$^{-2}$. For higher column densities, the models predict absorptions that are too deep for the ortho and para fundamental transitions (see Fig. \ref{pop_h218o}). For a visual extinction $A_V$$\sim$1-4 mag, the H$_2$O abundance in the photodesorption layer is 1.3 - 5.3 $\times$ 10$^{-7}$. This value is in very good agreement with the values predicted in photodesorption layers by the model of \citet{Hollenbach2009}, about 1.5 - 3 $\times$ 10$^{-7}$. 
The model predictions with this absorbing layer are shown in Fig. \ref{fig_h218o}. 
We cannot rule out a remnant contribution of the outflows in view of the profile of 548 GHz line, but this outflow contribution to the bulk of the emission is very likely negligible for the other transitions.

In the hot corino, among the o-H$_2^{18}$O transitions, only the 489 and 1181 GHz lines are optically thin, whereas among the p-H$_2^{18}$O transitions, only the 970, 1189, and 1606 GHz lines show an opacity lower than 1.
The emission of most of the transitions comes from both the hot corino and the outer envelope (see Fig.~\ref{pop_h218o}). Similarly to HDO, a rotational diagram analysis therefore cannot be considered for H$_2^{18}$O. 


\begin{figure}[!h]
\begin{center}
\includegraphics[scale=0.33]{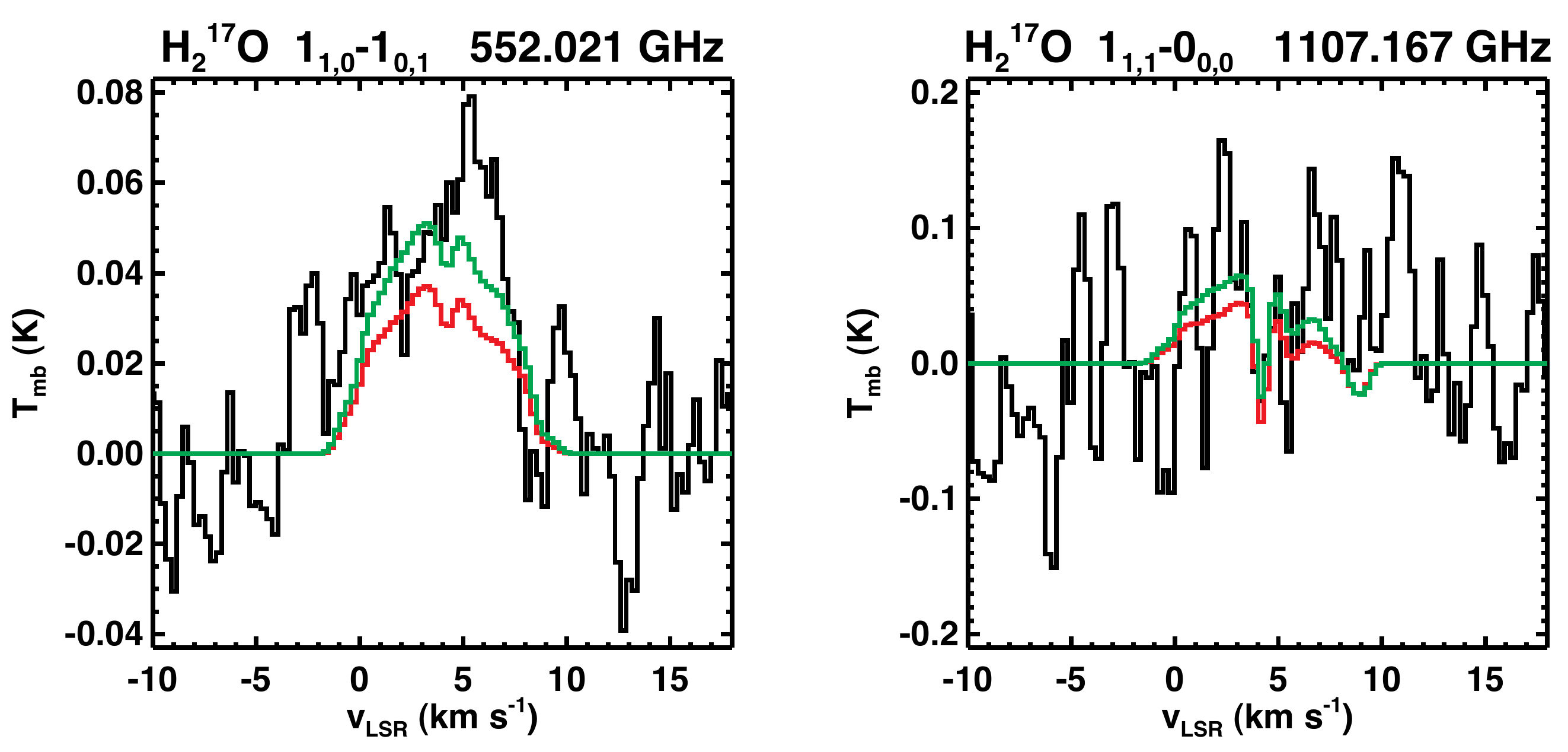}
\caption{In black: Fundamental H$_2^{17}$O lines observed with HIFI.
In red: model obtained with an inner H$_2^{17}$O abundance of 2.5 $\times$ 10$^{-9}$, an outer H$_2^{17}$O abundance of 7.5 $\times$ 10$^{-12}$, and an H$_2^{17}$O column density in the absorbing layer of  2.5 $\times$ 10$^{11}$ cm$^{-2}$.
In green: model obtained with an inner H$_2^{17}$O abundance of 5.0 $\times$ 10$^{-9}$, an outer H$_2^{17}$O abundance of 7.5 $\times$ 10$^{-12}$, and an H$_2^{17}$O column density in the absorbing layer of  2.5 $\times$ 10$^{11}$ cm$^{-2}$.}
\label{h217o}
\end{center}
\end{figure}

The ortho--H$_2^{17}$O 1$_{1,0}$-1$_{0,1}$ transition at 552 GHz has also been detected. To check the validity of the results, we ran a model similar to what has been done for H$_2^{18}$O, assuming a ratio $^{18}$O/$^{17}$O of 4 \citep{Wouterloot2008}. 
 Figure \ref{h217o} shows the predicted models for the 552 GHz line, as well as the para-H$_2^{17}$O 1$_{1,1}$-0$_{0,0}$ fundamental line undetected at 1107 GHz  for the best-fit of H$_2^{18}$O, i.e. for an inner H$_2^{17}$O abundance of 2.5 $\times$ 10$^{-9}$ and an outer H$_2^{17}$O  abundance of 7.5 $\times$ 10$^{-12}$.
A prediction with twice the inner abundance, situated in the 3$\sigma$ contour of H$_2^{18}$O, is also presented (Fig.~\ref{h217o}).
We see that the H$_2^{17}$O predictions agrees with the H$_2^{18}$O results, confirming the water abundances derived.

\section{Discussion}

The study of HDO in the deeply embedded low-mass protostar IRAS16293 had already been undertaken by \citet{Stark2004}, then by \citet{Parise2005}, leading to different results.
\citet{Stark2004} used a constant [HDO]/[H$_2$] abundance of 3 $\times$ 10$^{-10}$ throughout the envelope to fit the 1$_{0,1}$-0$_{0,0}$ transition at 465 GHz best, resulting in an HDO/H$_2$O ratio in the warm inner envelope of a few times 10$^{-4}$. However, their best fit cannot reproduce the deep absorption feature. Also, unlike our study, they do not succeed in reproducing the line width without introducing an outflow contribution; however, the model obtained with the outflow component fails to match simultaneously the line width and the peak intensity. This  could be explained by their source structure and the assumption of a constant abundance throughout the envelope. The dust temperature is restricted to a range from 12\,K to 115\,K. A structure with a lower inner radius (consequently an higher temperature) and an abundance rising in the inner part of the envelope may allow the line to become broader and reproduce the observation.
Moreover, the peak intensities predicted by their model for the lines at 226 and 241 GHz (0.001\,K and 0.006\,K, respectively) are clearly lower than what we observe here (0.30 and 0.32\,K, respectively). 
Afterwards, the results obtained by \citet{Stark2004} have been questioned by the study
of \citet{Parise2005}. Using IRAM and JCMT observations, they have found an enhancement of the abundance of HDO in the hot corino ($X_{\rm in}$ $\sim$ 1 $\times$ 10$^{-7}$) with respect to the outer envelope ($X_{\rm out}$ $\le$ 1 $\times$ 10$^{-9}$), in agreement with our results using both the collisional coefficients with He by \citet{Green1989} and with ortho and para-H$_2$ by \citet{Faure2011} (see Sect. 3.3). 
However, their results did not take the line profiles into account, specifically the deep absorption of the ground state transitions, and
only allowed a rather high upper limit of the outer abundance to be derived. 

\citet{Parise2005} conclude there is a jump in the HDO/H$_2$O ratio in the inner part of the envelope, using the H$_2$O abundance determined by \citet{Ceccarelli2000}.
Nevertheless, our results on the water abundance are quite different. Using H$_2^{16}$O data from ISO/LWS, highly diluted in a beam width of about 80$\arcsec$, \citet{Ceccarelli2000} obtained an outer abundance of 5 $\times$ 10$^{-7}$, while we found an outer abundance lower than 3.5 $\times$ 10$^{-8}$.
However, their derived inner abundance of $\sim$ 3 $\times$ 10$^{-6}$ is not so very different from our value, $X_{\rm in}$(H$_2$O) $\sim$ 5 $\times$ 10$^{-6}$.
The disagreement on determining the outer abundance can easily be explained by the fact that \citet{Ceccarelli2000} used H$_2^{16}$O data, contaminated by the outflows that must be added in the modeling.
The gas temperature was computed by assuming the water abundance derived by the ISO observations. However, since the gas is practically thermally coupled with the dust, the impact in the gas temperature is certainly negligible. The largest difference between the gas and dust temperature is less than 10\% \citep{Crimier2010}.
The discrepancy on the water abundance therefore results in a different deuteration ratio in the outer envelope. Considering the 3$\sigma$ uncertainty, the enhancement of the water deuterium fractionation in the hot corino cannot be confirmed in our study, in contrast to what was concluded in \citet{Parise2005}.
Table \ref{resume} summarizes the HDO/H$_2$O ratio determined by our analysis in the inner and outer envelope, as well as an estimation of this ratio in the absorbing layer.
With a 3$\sigma$ uncertainty, the water fractionation could be similar throughout the cloud with a HDO/H$_2$O ratio between 1.4\% and 5.8\% in the inner part and between 0.2\% and 2.2\% in the outer part. 
Also the HDO/H$_2$O ratio obtained here is in agreement with the upper limits of solid HDO/H$_2$O (from 0.5\% to 2\%) determined by observations of OH and OD stretch bands in four low-mass protostars \citep{Dartois2003,Parise2003}.

\begin{table*}[!ht]
\begin{center}
\caption{HDO/H$_2$O ratio}
\label{resume} 
\begin{tabular}{c |c |c| c| c |c }
\hline
\hline	
& \multicolumn{2}{|c|}{Hot corino} & \multicolumn{2}{c|}{ Outer envelope } & \multicolumn{1}{c}{ Photodesorption layer } \\
\cline{2-6}
& Best fit & 3$\sigma$ &  Best fit & 3$\sigma$ &  A$_V$ $\sim$ 1 - 4 mag\\
\hline
HDO & 1.7 $\times$ 10$^{-7}$ & 1.5 - 2.2 $\times$ 10$^{-7}$ &  8  $\times$ 10$^{-11}$ & 4.6 - 10.0 $\times$ 10$^{-11}$ & $\sim$ 0.6 - 2.4  $\times$ 10$^{-8}$ \\
H$_2$O & 5 $\times$ 10$^{-6}$ & 3.8 - 10.5 $\times$ 10$^{-6 }$& 1.5 $\times$ 10$^{-8}$ & 4.5 - 24.5 $\times$ 10$^{-9}$ & $\sim$ 1.3 - 5.3 $\times$ 10$^{-7}$ \\
HDO/H$_2$O & 3.4\% & 1.4\% - 5.8\% & 0.5\% & 0.2\% - 2.2\% & $\sim$ 4.8\%$^a$\\
\hline
\end{tabular}
\end{center}
$^{a}$ This ratio remains valid if the absorbing layer is not due to the photodesorption, but is the result of a freeze-out timescale longer than the protostellar age at low densities.
\end{table*}

A consequential result of this paper is the similarity of the HDO/H$_2$O ratio derived in the hot corino ($\sim$ 1.4-5.8\%) and in the added outer absorbing layer ($\sim$ 4.8\%).
Indeed, the ratio is within the same order of magnitude, although the densities of these two regions are considerably different.
The density in the hot corino is a few times 10$^{8}$ cm$^{-3}$. In the absorbing layer, it is about 10$^3$-10$^5$ cm$^{-3}$.
Consequently, water shows a different behavior from other molecules such as methanol and formaldehyde, that need CO ices to be formed.
In particular, \citet{Bacmann2003,Bacmann2007} show that the deuteration of H$_2$CO and CH$_3$OH increases with the CO depletion in starless dense cores, increasing itself with the H$_2$ density \citep{Bacmann2002}. 
The H$_2$CO and CH$_3$OH deuteration is therefore sensitive to the density of the medium in which they form.
In contrast, the HDO/H$_2$O ratio does not show any difference for different H$_2$ densities. 
This would therefore mean that water has formed before the collapse of the protostar and that the HDO/H$_2$O ratio has been preserved during the gravitational collapse. The deuteration fractionation of water would therefore remain similar, both in the inner region of the protostar, where the density has strongly increased, and in the outer region that has not been affected by the collapse. 
Another argument emphasizes this hypothesis. To obtain such high deuteration ratios of HDCO/H$_2$CO \citep[$\sim$15\%;][]{Loinard2001} and CH$_2$DOH/CH$_3$OH \citep[$\sim$30\%;][]{Parise2004} in IRAS16293 compared with HDO/H$_2$O$\sim$3\%, the density at which the molecules form should be higher for the H$_2$CO and CH$_3$OH formation than for water formation. 
The collapse should already have started to allow the CO molecules to freeze-out and form formaldehyde and methanol at the grain surface.
On the contrary, water would form at low densities in the early stages of the star formation before the protostellar collapse, as suggested by \citet{Dartois2003} and \citet{Parise2003}. 
This is also consistent with the fact that H$_2$O ices appear at relatively low extinction in the direction of the Taurus cloud \citep[e.g.][]{Jones1984}.
Recently, similar conclusions have been mentioned in \citet{Cazaux2011}. Using a grain surface chemistry model, they show that the deuteration of formaldehyde is sensitive to the gas D/H ratio as the cloud undergoes gravitational collapse, while the HDO/H$_2$O ratio is constant as the cloud collapses and is set during the formation of ices in the translucent cloud.

The HDO/H$_2$O ratio in the low-mass protostar IRAS16293 is close to what is found in the low-mass protostar NGC1333-IRAS2A, as determined by \citet{Liu2011}: higher than 1\% in the hot corino and between 0.9\% and 18\% at 3$\sigma$. However, this ratio of a few percent does not seem typical of all the Class 0 protostars since an upper limit of 6 $\times$ 10$^{-4}$ has been determined in the inner part of the envelope of the low-mass protostar NGC1333-IRAS4B \citep{Jorgensen2010}. 
The determination of the water deuterium fractionation in a larger sample of Class 0 protostars  would allow us to know, from a statistical point of view, whether the HDO/H$_2$O ratio is rather about 10$^{-4}$ -- 10$^{-3}$ as observed in the NGC1333-IRAS4B protostar, in comets \citep[$\sim$ 3 $\times$ 10$^{-4}$; e.g. ][]{Bockelee1998} and in the Earth's oceans \citep[$\sim$ 1.5 $\times$ 10$^{-4}$;][]{Lecuyer1998} or rather about a few percent, like the young stellar objects NGC1333-IRAS2A and IRAS16293. 
With a value of about 10$^{-4}$ in the protostellar phase, the fractionation ratio could be conserved throughout the different stages of the star formation until the formation of the the solar-system objects. Higher values in the protostellar phase invoke mechanisms in the gas phase and/or on the grain surfaces to explain the decrease in the deuterium fractionation for water from the protostar formation to the comets and solar system formation.
The determination of the HDO/H$_2$O ratio in a larger sample would allow us to understand whether the deuteration of water in protostars is somewhat similar to the solar system value and whether the environment or the conditions for the prestellar core phase play a role.

Finally, we can estimate for the first time the D$_2$O/H$_2$O ratio in a low-mass protostar.
According to the results of \citet{Vastel2010}, the column density of D$_2$O in the cold envelope ($T <$ 30\,K) is 1.65 $\pm$ 1.41 $\times$ 10$^{12}$ cm$^{-2}$ at a 3$\sigma$ uncertainty. 
If we consider that the absorption of the two D$_2$O fundamental lines can be both due to the cold envelope ($T <$ 30K), as well as to the absorbing layer, the D$_2$O/H$_2$O ratio is on average about 10$^{-3}$ and in the interval 1.1 $\times$ 10$^{-4}$ - 3.75 $\times$ 10$^{-3}$ at 3$\sigma$. As for the D$_2$O/HDO ratio, it is about 6\% and between 0.8\% and 11.6\% at 3$\sigma$.
The HDO/H$_2$O and D$_2$O/H$_2$O ratio follow the statistical distribution (even considering the 3$\sigma$ uncertainty) as determined in \citet{Butner2007}:
\begin{equation}
\rm  \left( \frac{D_2O}{HDO} \right)_{grain} \ge  \frac{1}{4} \left(\frac{HDO}{H_2O}\right)_{grain}.
\end{equation}
These estimations of deuteration ratios in a low-mass protostar will certainly provide better constraints on the gas-grain chemistry models. Indeed, HDO is trapped on the grain surfaces before the thermal desorption due to the heating from the accreting protostar or before its photodesorption at the edges of the molecular cloud.

\section{Conclusion}

This study is the first one to use such a large number of lines to model deuterated water. 
Thanks to the numerous HDO transitions observed with Herschel/HIFI and four other lines observed with ground-based telescopes, we have succeeded in accurately determining the abundance of HDO throughout the envelope of the protostar and particularly the outer abundance that \citet{Parise2005} could not constrain. 
To estimate the abundances, new collisional coefficients computed with ortho and para-H$_2$ by \citet{Faure2011} and for a wide range of temperatures were used. The best-fit inner abundance $X_{\rm in}$(HDO) is about 1.7 $\times$ 10$^{-7}$, whereas the best-fit outer abundance $X_{\rm out}$(HDO) is about 8 $\times$ 10$^{-11}$.
To model the deep HDO absorption lines, it has been necessary to add an outer absorbing layer with an HDO column density of 2.3 $\times$ 10$^{13}$ cm$^{-2}$. 
In addition, detections of several H$_2^{18}$O transitions, as well as small upper limits on other transitions, have allowed us to constrain both the inner and outer water abundances. Assuming a standard isotopic ratio H$_2^{18}$O/H$_2^{16}$O~=~500, the inner abundance of water is about 5~$\times$~10$^{-6}$ and the outer abundance about 1.5~$\times$~10$^{-8}$. The water column density in the added absorbing layer is about 5~$\times$~10$^{14}$~cm$^{-2}$. If we consider that this absorbing layer is created by the photodesorption of the ices at an $A_V$ of $\sim$1-4 mag, the water abundance is about 1.5~-~3~$\times$~10$^{-7}$ nicely consistent with the values predicted by \citet{Hollenbach2009}.
The deuterium fractionation of water is therefore about 1.4-5.8\% in the hot corino, 0.2-2.2\% in the colder envelope, and 4.8\% in the added absorbing layer. The 3$\sigma$ uncertainties determined in both the inner and the outer part of the envelope are small due to both the well constrained HDO and H$_2^{18}$O abundances (see Table 2). 
These results do not permit any conclusion on an enhancement of the fractionation ratio in the inner envelope with respect to the outer envelope.
The similar ratios derived in the hot corino and in the absorbing layer suggest that water forms before the gravitational collapse of the protostar, unlike formaldehyde and methanol, which form later after the CO molecules have depleted on the grains.
In the cold envelope ($T <$ 30K), the D$_2$O/HDO ratio is estimated with a value of 6\%.
The HDO/H$_2$O ratios found here are clearly higher than those observed in comets ($\sim$ 0.02\%). The water deuterium fractionation ratio has to be estimated in more low-mass protostars to determine if IRAS16293 is an exception among the Class 0 sources or a typical protostar that could lead to the formation of a planetary system similar to our Solar System.
In the latter case, processes should be at work to reprocess the water deuteration ratio before the cometary formation stage.

The strong constraints obtained here emphasize the necessity to observe several lines in a broad frequency and energy range, as done with the HIFI spectral survey, to precisely estimate the water deuterium fractionation. In particular, the HDO 1$_{1,1}$-0$_{0,0}$ fundamental line at 894 GHz is a key line in the modeling because it shows emission both from the hot corino and the outer part of the envelope. In addition, both this transition and the 1$_{0,1}$-0$_{0,0}$ transition at 465 GHz present a deep absorption probing the absorbing layer. The HDO 3$_{1,2}$-2$_{2,1}$ and 2$_{1,1}$-2$_{1,2}$ transitions observed at 226 and 242 GHz, respectively, with the IRAM-30m telescope also bring useful information as they entirely probe the hot corino.
These transitions as well as several H$_2^{18}$O transitions lie in the ALMA\footnote{Atacama Large Millimeter Array} spectral range that will hopefully allow the HDO/H$_2$O ratio to be studied and constrained in many low-mass protostars with very high spatial resolution.

\section*{Acknowledgements}

HIFI was designed and built by a consortium of institutes and university departments from across Europe, Canada, and the United States under the leadership of SRON Netherlands Institute for Space Research, Groningen, The Netherlands, with major contributions from Germany, France and the US. Consortium members are: Canada: CSA, U.Waterloo; France: CESR, LAB, LERMA, IRAM; Germany: KOSMA, MPIfR, MPS; Ireland, NUI Maynooth; Italy: ASI, IFSI-INAF, Osservatorio Astrofisico di Arcetri-INAF; Netherlands: SRON, TUD; Poland: CAMK, CBK; Spain: Observatorio Astron\'{o}mico Nacional (IGN), Centro de Astrobiolog\'{i}a (CSIC-INTA). Sweden: Chalmers University of Technology - MC2, RSS \& GARD; Onsala Space Observatory; Swedish National Space Board, Stockholm University Ð Stockholm Observatory; Switzerland: ETH Zurich, FHNW; USA: Caltech, JPL, NHSC. We thank many funding agencies for financial support.
LW thanks the COST `Chemical Cosmos' program, as well as the CNRS national program `Physique et Chimie du Milieu Interstellaire' for partial support. 
The collision coefficients calculations presented in this paper were performed at the Service Commun de Calcul Intensif de l'Observatoire de Grenoble (SCCI).

\bibliographystyle{aa}
\bibliography{biblio_i16293_hdo}

\end{document}